\documentclass[nojss]{jss}

\usepackage{thumbpdf,lmodern, expl3}

\usepackage{framed}
\usepackage{fancyvrb}
\usepackage{amsmath}
\usepackage{multirow,listings,subcaption}
\usepackage{tcolorbox,amsmath,amssymb}
\usepackage[english]{babel}

\DefineVerbatimEnvironment{CodeInput}{Verbatim}{fontsize=\small}

\author{Gordon J. Ross\\University of Edinburgh\\gordon.ross@ed.ac.uk \And
Dean Markwick\\University College London \And Priyanshu Tiwari \\ IIT Kanpur}
\Plainauthor{Gordon J. Ross, Dean Markwick, Priyanshu Tiwari}

\title{\pkg{dirichletprocess}: An \proglang{R} Package for Fitting Complex Bayesian Nonparametric Models}
\Plaintitle{dirichletprocess: An R Package for Fitting Complex Bayesian Nonparametric Models}
\Shorttitle{The \pkg{dirichletprocess} Package}

\Abstract{
The \pkg{dirichletprocess} package provides software for creating flexible Dirichlet process objects. Users can perform nonparametric Bayesian analysis using Dirichlet processes without the need to program their own inference algorithms. Instead, the user can utilise our pre-built models or specify their own models whilst allowing the \pkg{dirichletprocess} package to handle the Markov chain Monte Carlo sampling. Our Dirichlet process objects can act as building blocks for a variety of statistical models including: density estimation, clustering and prior distributions in hierarchical models.

}

\Plainkeywords{dirichlet process, bayesian inference, nonparametric statistics, R}

\Address{
  Gordon J. Ross\\
  School of Mathematics\\
  University of Edinburgh\\
  James Clerk Maxwell Building \\
  Edinburgh, UK\\
  E-mail: \email{gordon.ross@ed.ac.uk}\\
  URL: \url{http://gordonjross.co.uk/} \\
  \emph{and} \\
  Dean Markwick \\
  Department of Statistical Science\\
  University College London\\
  1-19 Torrington Place \\
  London, UK\\
  E-mail: \email{dean.markwick@talk21.com}\\
  URL: \url{http://dm13450.github.io/} \\
  \emph{and} \\
  Priyanshu Tiwari
  Indian Institute of Technology Kanpur
  Kalyanpur
  Kanpur -208 016
  Email: \url{tiwari.priyanshu.iitk@gmail.com}
}

\begin{document}

\section[Introduction]{Introduction} \label{sec:intro}

Many applications of statistical models require parametric specifications of the probability distributions of interest -- that observations are generated by Normal distributions, that group-level fixed effects can be modeled by Gamma distributions, and so on. However in some situations there will be insufficient prior information and data to justify such parametric assumptions. In this case, nonparametric methods allow a more flexible and robust specification of distributions, so their  essential features can be `learned from the data' rather than specified in advance.

While frequentist nonparametrics has a long history, Bayesian nonparametrics was a relatively dormant field until the mid 1990s. Although much of the theory of nonparametric priors had been worked out in previous decades \citep{ferguson_bayesian_1973}, computational issues prevented their widespread adoption. This changed with the development of posterior simulation methods such as Metropolis-Hastings \citep{hastings_monte_1970} and the Gibbs sampler \citep{geman_stochastic_1984} which were first applied to the task of nonparametric density estimation using Dirichlet process (DP) mixtures in a seminal paper by \cite{escobar_bayesian_1995}. This kickstarted research into Bayesian nonparametrics, which has now become one of the most popular research areas in the statistics and machine learning literature. While there are now several widely used models within the field of Bayesian nonparametrics including the Gaussian process, Beta process and Polya trees, the Dirichlet process mixture model (DPMM) has been the most popular due to its wide applicability and elegant computational structure.

At the most basic level, the DPMM can be viewed as an infinite dimensional mixture model which represents an unknown density $f(y)$ as:

\begin{equation*}
f(y) = \int k(y \mid \theta) p(\theta \mid G) \mathrm{d} \theta, \quad G \sim \text{DP}(\alpha, G_0),
\end{equation*}
where $k(\cdot \mid \theta)$ denotes the mixture kernel, and the mixing distribution $G$ is assigned a nonparametric Dirichlet process prior with a base measure $G_0$ and concentration parameter $\alpha$. In the most widely used DPMM, the mixture kernel is taken to be Gaussian so that $\theta = (\mu,\sigma^2)$ and $k(y\mid\theta) = N(y\mid \mu,\sigma^2)$ with a conjugate Normal Inverse-Gamma specification for $G_0$. The infinite dimensional nature of such a model makes it capable of approximating any continuous distribution to an arbitrary degree of accuracy.

The use of DPMMs is not restricted to simply estimating the density of observed data. Instead, the DPMM can be used at any level in a hierarchical model where it is considered necessary to represent a density nonparametrically due to a lack of knowledge about its parametric form. For example, consider a (multilevel) random effects model where there are $J$ groups of observations, with observations in each group $j$ following a Gaussian distribution with a group-specific mean $\mu_j$ and a common variance $\sigma^2$. To share information across groups, the means $\mu_j$ are assumed to be exchangeable and assigned a prior $p(\mu_j)$. If $y_{i,j}$ denotes the $i^{th}$ observation in group $j$, then the model is:
\begin{align*}
y_{i,j} & \sim N(y_{i,j} \mid \mu_j,\sigma^2), \\
\mu_j & \sim p(\mu_j \mid \gamma),
\end{align*}
where $\gamma$ is the (hyper-)parameters of the prior. This model is an example of \textbf{partial pooling}, where the inference for each mean $\mu_j$ is based on the means of each of the other $J-1$ groups, allowing information to be shared across groups. As a biased estimation technique, it is related to methods such as the Lasso estimator, and usually results in more accurate estimation than estimating each group mean separately. However, completing the model specification requires choosing a form for the prior distribution of group means $p(\mu_j \mid \gamma)$, which is made more difficult since the group means may not be observable with a high degree of accuracy, particularly when the number of observations is small. In this case, using a nonparametric DPMM specification for $p(\mu_j \mid \gamma) $ would avoid the risks of potentially specifying an inappropriate parametric form.

Typically when working with DPMMs, the posterior distributions are analytically intractable, so inference instead usually involves computational simulation. A variety of simulation algorithms based around Gibbs sampling and Metropolis-Hastings have been developed to draw samples from DPMM posterior distributions, which can then be used for inference. As such, the widespread adoption of DPMMs has to some extent been held back by the level of statistical and programming literacy required to implement and use them. %The purpose of the \pkg{dirichletprocess} package is to provide a unified implementation of these simulation algorithms for a very wide class of DP mixture models which makes it easy to incorporate DPMMs into hierarchical models.

The purpose of the \pkg{dirichletprocess} package is not to automate a pre-specified range of tasks, but instead to represent Dirichlet process mixture models as objects in \proglang{R} so that they can be used as building blocks inside user-specified models. The target audience is hence users who are working with models which use a DPMM at some stage, and who require a DPMM implementation which can be used as part of a more general estimation scheme. As such, the number of tasks which can be achieved using \pkg{dirichletprocess} is quite large, with an emphasis on flexibility and extensibility rather than a fixed collection of model-specific routines.

Several software packages are available for Dirichlet process and related Bayesian nonparametric mixture modelling. Historically, \pkg{DPpackage} \citep{Jara2011DPpackage} provided an important collection of implementations for specific Dirichlet process models, although it is now archived on CRAN. More recent \proglang{R} packages such as \pkg{PReMiuM} \citep{Liverani2015PReMiuM} and \pkg{BNPmix} \citep{Corradin2021BNPmix} focus on more specialised model classes, including profile regression, clustering, density estimation, and dependent or partially exchangeable extensions. More general probabilistic programming frameworks such as Stan \citep{Carpenter2017Stan} and PyMC \citep{pymc2023} are broader in scope, but are not built around the object-based Dirichlet process mixture workflow used here; in particular, Stan does not directly sample latent discrete parameters and instead requires such models to be reformulated by marginalisation. In contrast, the aim of \pkg{dirichletprocess} is to provide a flexible object-based framework for constructing, fitting, and extending Dirichlet process mixture models within \proglang{R}, rather than a fixed collection of model-specific procedures. The corresponding trade-off is that packages designed for more specialised model classes or broader probabilistic programming tasks may be preferable when those settings align closely with the application of interest.

%The design philosophy of the \pkg{dirichletprocess} package is the polar opposite of the existing  \pkg{DPpackage} \proglang{R} package which also provides an implementation of DPMMs. The purpose of \pkg{DPpackage} is to give a fast implementation of several common tasks where DPs are used, such as density estimation using Gaussian mixtures, and a variety of specific regression models. While the \pkg{DPpackage} package is very useful in these situations, it cannot be used for any applications of DPs which do not fall into one of the pre-specified tasks incorporated in the package, or use mixture kernels or base measures other than those provided.

%In contrast, the purpose of the \pkg{dirichletprocess} package is not to automate a pre-specified range of tasks, but instead to represent DPMMs as objects in \proglang{R} so that they can be used as building blocks inside user-specified  models. The target audience is hence users who are working with  (possibly hierarchical) models which uses a DPMM at some stage, and who hence require a DPMM implementation which can be used as a part of a more general model estimation scheme.  As such, the number of tasks which can be achieved using the \pkg{dirichletprocess} is quite large, although the trade-off is that the functions in this package will be slower than those in \pkg{DPpackage} when it comes to the specific models which it implements.

Key features of the \pkg{dirichletprocess} package include:
\begin{itemize}
\item An implementation of DP mixture models for various types of mixture kernel including the Gaussian, Beta, Multivariate Normal and Weibull.
\item Implementation of DP posterior sampling algorithms in both the conjugate and non-conjugate cases.
\item An object-based interface which allows the user to work directly with DP objects in \proglang{R} so that they can be incorporated into hierarchical models.
\end{itemize}

\subsection{A Technical Note}

The ultimate purpose of this package is to represent Dirichlet process mixture models as
objects in \proglang{R}, so that they can be manipulated and used as building blocks within
larger statistical workflows. At the user level, the package therefore uses an S3 object-based
interface, since this makes it straightforward to construct, inspect, and extend DP mixture
objects directly in \proglang{R}.

At the same time, the computationally intensive parts of the package are not handled purely
in \proglang{R}. For the mixture models discussed in this paper, including the Gaussian,
multivariate Gaussian, Beta, and Weibull cases, the core sampling routines are implemented
in \proglang{C++}. In particular, the repeated MCMC transition steps used when fitting these
models are carried out by compiled code, with the \proglang{R} layer providing the object
representation, model specification, and user-facing workflow.

This separation reflects the design philosophy of the package. The \proglang{R} interface is
intended to make it easy for users to work with Dirichlet process mixture models as objects,
to embed them within broader estimation schemes, and to define new mixture specifications
when needed. The compiled implementation is intended to ensure that the standard models
provided by the package can be fitted without relying on a purely interpreted sampler.

As such, the package aims to combine two features which are often in tension: a flexible and
extensible object-based interface in \proglang{R}, and compiled implementations of the core
sampling routines for the main mixture models discussed in this paper.

%Current alternatives for nonparametric inference include Stan, PyMC3 and Edward. However, whilst all three packages are much more general than the \pkg{dirichletprocess} offerings, they do not offer ease of customisation that \pkg{dirichletprocess} does. Firstly, Stan \citep{carpenter_bob_stan:_2016}, does not allow you to specify discrete parameters in models. As Dirichlet process models require cluster labels which are inherently discrete parameters you are unable to build Dirichlet process models directly in Stan. For both the Python libraries Edward and PyMC3, examples exist of building Dirichlet process models. However, these are built on top of TensorFlow and Theano \citep{tran_edward:_2016, salvatier_probabilistic_2016}, therefore, being able to build Dirichlet process objects into statistical work flows would require learning these external libraries. Instead our package \pkg{dirichletprocess} is written natively in R and abstracts the difficulties away, allowing users to write Dirichlet process models in R code and not worry about computational details.

\section{Background information}
\label{sec:background}

%need to include the fact that the posterior is also a DP
This section provides background information about the Dirichlet Process, and includes the key mathematical properties around which the sampling algorithms in the \pkg{dirichletprocess} package are based. It can be skipped by those who simply want to know how to use the package, which is discussed in Section  \ref{sec:package}.

It is commonly required to learn the unknown probability distribution $F$ which generated the observations $y_1,\ldots,y_n$. In parametric Bayesian inference, $F$ is assumed to belong to a known family of distributions (such as the Normal or Exponential) with an associated parameter vector $\theta$ of finite length  which must be estimated. This leads to the model:
\begin{align*}
y_i \sim F(y_i \mid \theta), \\
\theta \sim p(\theta \mid \gamma),
\end{align*}
where $p(\theta \mid \gamma)$ denotes the prior distribution and $\gamma$ are the prior parameters. The task of inference then involves finding an appropriate value for $\theta$, which is equivalent to choosing which member of the specified family of distributions gives the best fit to the data.

In practice however, it may not be clear how to choose an appropriate parametric family of distributions for $F$. If the wrong family is chosen, then conclusions based on the estimated model may be highly misleading. For example, if it is assumed that $F$ has a Normal distribution with unknown parameters $\theta=(\mu,\sigma^2)$ when in fact the true $F$ is heavy-tailed, this can lead to severe underestimation of the probability of extreme events occurring \citep{coles_introduction_2001}.

This problem can be avoided by using a nonparametric prior specification which puts positive prior mass on the whole space of probability densities rather than on a subspace spanned by the finite-length parameter vector $\theta$. This allows the estimated $F$ to adapt to the data, rather than being restricted to a particular family of distributions such as the Normal or Exponential. The Dirichlet process (DP) is one of the most widely used Bayesian nonparametric priors, due to its flexibility and computational simplicity. Our aim in this section is not to give a full treatment of the Dirichlet processes, and a reader unfamiliar with them should refer to a standard reference such as \cite{escobar_bayesian_1995}. Instead we will only focus on describing the properties of the DP that are directly relevant to their implementation in the \pkg{dirichletprocess} package.

The basic DP model has the form:
\begin{align*}
y_i & \sim F, \\
F & \sim \text{DP} (\alpha, G_0),
\end{align*}
where $G_0$ is known as the \textbf{base measure} and encapsulates any prior knowledge that might be known about $F$.  Specifically, it can be shown that $\mathbb{E}[F \mid G_0,\alpha] = G_0$. The concentration parameter $\alpha$ specifies the prior variance and controls the relative contribution that the prior and data make to the posterior, as the following result shows.

\begin{tcolorbox}

\textbf{Key Property 1}: The DP is a conjugate prior in the following sense: if $y_1,\ldots,y_n \sim F$ and $F \sim \text{DP} (\alpha, G_0)$, then:

$$F \mid y_1,\ldots,y_n \sim \text{DP} \left( \alpha + n, \frac{\alpha G_0 + \sum_{i=1}^n \delta_{y_i}}{\alpha+n}\right),$$
where $\delta_{y_i}$ denotes a point-mass at $y_i$. In other words, the posterior distribution of $F$ is a weighted sum of the base measure $G_0$ and the empirical distribution of the data, with the weighting controlled by $\alpha$.

\end{tcolorbox}

The DP is a prior distribution over the space of probability distributions. As such, samples from a DP are probability distributions. The stick-breaking representation first introduced by \cite{sethuraman_constructive_1994} shows what such samples look like:

\begin{tcolorbox}

\textbf{Key Property 2}:
Suppose that $F \sim \text{DP} (\alpha, H)$ is a random probability distribution sampled from a DP prior. Then with probability 1, F can be written as:

$$F = \sum_{k=1}^\infty w_k \delta_{\phi_k}, \quad \phi_k \sim G_0$$
where
$$w_k = z_k \prod_{i=1}^{k-1} (1-z_i),\quad z_i \sim \text{Beta}(1,\alpha).$$

\end{tcolorbox}

In other words, random probability distributions can be sampled from a DP by first drawing a collection of samples ${z_i}$ from a Beta distribution, transforming these to produce the weights $\{w_i\}$, and then drawing the associated atoms from $G_0$. Note that in order for $F$ to be a true draw from a $\text{DP}$, an infinite number of such weights and atoms must be drawn. However in practice, the above summation can be truncated with only a finite number $N$ of draws, while still providing a very good approximation.

By combining Key Properties 1 and 2, we can sample a DP from its posterior distribution $F \mid y_1,\ldots,y_n$ as follows:

\begin{tcolorbox}

\textbf{Key Property 3}:  If $y_1,\ldots,y_n \sim F$ and $F \sim \text{DP} (\alpha, G_0)$ then we can draw a (truncated) sample probability distribution from the posterior $F\mid y_1,\ldots,y_n$ as follows:

$$F = \sum_{k=1}^N w_k \delta_{\phi_k}, \quad \phi_k \sim \frac{\alpha G_0 + \sum_{i=1}^n \delta_{y_i}}{\alpha+n},$$
where
$$w_k = z_k \prod_{i=1}^{k-1} (1-z_i),\quad z_i \sim \text{Beta}(1,\alpha+n).$$
and $N$ is a truncation parameter.

\end{tcolorbox}

%The final key property of the DP concerns the marginal distribution  $p(y_1,\ldots,y_n \mid G_0,\alpha)$, with $F$ integrated out. SHOULD THIS BE HERE OR IN MIXTURE SECTION?
%\textbf{Key Property 4:}

\subsection{Dirichlet Process Mixtures} \label{subsec:dpm}
The stick breaking representation in Key Property 2 above shows that  probability distributions sampled from a DP are discrete with probability 1. As such, the DP is not an appropriate prior for $F$ when $F$ is continuous. Therefore, it is usual to adopt the following mixture specification instead, which is known as the Dirichlet process mixture model (DPMM):
\begin{align}
\begin{split}
y_i &\sim k(y_i \mid \theta_i), \\
\theta_i &\sim F, \\
F &\sim \text{DP} (\alpha, G_0).
\end{split}
\label{eqn:dpmm}
\end{align}
In other words, $F$ has a DP prior as before, but rather than the data $y_i$ being drawn from $F$, it is instead the mixture parameters $\theta$ which are draws from F.  These $\theta$ values then act as the parameters of a parametric kernel function $k(\cdot)$, which is usually continuous. The most commonly used example is the Gaussian mixture model where $\theta_i = (\mu_i,\sigma^2_i)$ so that $k(y_i \mid \theta_i) = N(y_i \mid \mu_i,\sigma^2_i)$.

The key point here is that since $F$ is discrete, two independent draws $\theta_i$ and $\theta_j$ from $F$  can have identical values with a non-zero probability. As such, the DPMM can be seen as sorting the data into clusters, corresponding to the mixture components. The above model can hence be written equivalently as the following mixture model, which is infinite dimensional and can hence be viewed as a generalization of the finite mixture models commonly used in nonparametric statistics:
\begin{align*}
\begin{split}
y_i &\sim G, \\
G & = \int k(y_i \mid \theta) F(\theta) \mathrm{d} \theta ,\\
F &\sim \text{DP} (\alpha, G_0).
\end{split}
\label{eqn:dp2}
\end{align*}
When the DPMM is used in practice, interest may focus on several different parameters of the posterior distribution. In some cases, the primary object of interest will be all $n$ of the ${\theta_i}$ parameters from Equation \ref{eqn:dpmm} which are associated with the $n$ observations. This is particularly the case in clustering applications, where the goal is to assign similar observations to the same cluster (i.e. to identical values of $\theta$). However in other situations it will be the distribution $F$ which is of primary interest. The \pkg{dirichletprocess} package returns posterior samples of all these quantities, so that the user can decide which are most relevant.

Posterior inference in the \pkg{dirichletprocess} package is based around the Chinese Restaurant Process sampler \citep{neal_markov_2000}. This is a Gibbs-style algorithm based on the DPMM representation in Equation \ref{eqn:dpmm} above, and draws samples of $\theta_1,\ldots,\theta_n$ from their posterior with the distribution $F$ integrated out.

\begin{tcolorbox}
\textbf{Key Property 4:}   Let $\theta_{-i}$ denote the set of $\theta$ values with $\theta_i$ excluded, i.e.\ $\theta_{-i} =
(\theta_1, \ldots, \theta_{i-1}, \theta_{i+1}, \ldots, \theta_n)$. Then the posterior distribution for $\theta_i$ conditional on the
other model parameters is:
\[
p(\theta_i \mid \theta_{-i}, y_{1:n}, \alpha, G_0) = \sum_{j \neq i} q_{i,j}\delta(\theta_j) + r_i H_i,
\]
\[
q_{i,j} = b k(y_i, \theta_j), \qquad r_i = b \alpha \int k(y_i, \theta)\,dG_0(\theta)
\]
where $b$ is set such that $\sum_{j \neq i} q_{i,j} + r_i = 1$ and $H_i$ is the posterior distribution of $\theta$
based off of the prior base measure $G_0$.
\end{tcolorbox}

Based on this result, Gibbs sampling is used to repeatedly draw each value of $\theta_i$ in turn from
its posterior distribution, with all other variables held constant. An important distinction
needs to be made between the conjugate case where the $G_0$ base measure is the conjugate
prior for $\theta$ with respect to the kernel $k(\cdot)$, and the non-conjugate case where there is not a closed
form for the posterior distribution. In the conjugate case, the integral in Key Property 4 can
be computed analytically and the resulting distribution is simply the predictive distribution.
In this case, the $\theta_i$ values can be sampled directly from their true posterior distribution.

In the non-conjugate case things are slightly more difficult, and the integral in Key Property 4
cannot be evaluated. As such, numerical techniques must be used instead, which will typically
result in slower computation. The \pkg{dirichletprocess} package handles the non-conjugate case
by using ``Algorithm 8'' from \citet{neal_markov_2000}, which is one of the most widely used techniques
for performing this sampling.

In both the conjugate and non-conjugate cases, the Gibbs sampling is conceptually similar,
with the new values of $\theta_i$ being proposed sequentially from their respective posterior distributions.
However in practice, this can result in poor mixing of the Gibbs sampler. One
approach to speed up convergence is to add in an additional update of the $\theta_i$ values at the end of
the sampling. For each cluster and its associated data points we update the cluster parameter
using the posterior distribution
\begin{equation}
p(\phi \mid y_{j:\theta_j=\phi}) \propto G_0(\phi)\prod_{j:\theta_j=\phi} k(y_j \mid \phi).
\label{eq:clusterposterior}
\end{equation}
For a conjugate base measure, this posterior distribution is tractable and thus can be sampled
directly. For a non-conjugate $G_0$, a posterior sample is obtained using the Metropolis-Hastings
algorithm \citep{hastings_monte_1970}.

For simple density estimation and non-hierarchical predictive tasks, having a posterior sample
of $\theta_{1:n}$ will be sufficient for inference, and the distribution $F$ is not of any intrinsic interest.
However when the DP is used as part of a hierarchical model such as in the regression and
point process examples we discuss later, it is also necessary to have samples from the posterior
distribution of $F$. These can be obtained using the following property:

\begin{tcolorbox}
\textbf{Key Property 5}: Given the model from Eq.\ \eqref{eqn:dpmm}, let $\theta_1, \ldots, \theta_n$ be a sample from
the posterior $p(\theta_{1:n} \mid y_{1:n}, \alpha, G_0)$ drawn using the CRP sampler. Then
\[
p(F \mid \theta_{1:n}, y_{1:n}, \alpha, G_0) = p(F \mid \theta_{1:n}, \alpha, G_0)
\]
is conditionally independent of $y_{1:n}$. As such,
$\theta_{1:n}$ can be considered as an i.i.d.\ sample from $F$, and so $F$ can be sampled from its
posterior distribution using Key Property 3 above:
\[
F = \sum_{k=1}^N w_k \delta_{\phi_k}, \qquad
\phi_k \sim \frac{\alpha G_0 + \sum_{i=1}^n \delta_{\theta_i}}{\alpha + n},
\]
where
\[
w_k = z_k \prod_{i=1}^{k-1}(1-z_i), \qquad z_k \sim \text{Beta}(1, \alpha + n),
\]
and $N$ is again a truncation parameter.
\end{tcolorbox}

\subsection{Hyperparameter Inference} \label{subsec:hyperparam}

In the discussion above, we treated the concentration parameter $\alpha$ and the base measure $G_0$ as fixed. In practice, it can be useful to learn one or both of these from the data rather than specifying them in advance. In the \pkg{dirichletprocess} package, such updates are implemented on a model-specific basis.

\subsubsection{Inferring the Concentration Parameter}\label{subsubsec:concentration}

Following \cite{west_hyperparameter_1992}, we place a $\text{Gamma}(a,b)$ prior on $\alpha$, using the shape--rate parameterisation. The corresponding posterior distribution depends only on the number of distinct values among $\theta_{1:n}$. More specifically, given the model in Equation \eqref{eqn:dpmm}, let $\theta_1,\ldots,\theta_n$ denote a sample from the posterior $p(\theta_{1:n} \mid y_{1:n}, \alpha, G_0)$, and suppose that there are $k$ distinct values in this sample. A draw from $p(\alpha \mid \theta_{1:n}, y_{1:n}, G_0)$ can then be obtained using the following auxiliary-variable scheme:

\begin{itemize}
\item Draw $z \sim \text{Beta}(\alpha + 1, n)$.
\item Define
\[
\tilde{\pi}_1 = a + k - 1, \qquad \tilde{\pi}_2 = n\bigl(b - \log z\bigr),
\]
and hence
\[
\pi = \frac{\tilde{\pi}_1}{\tilde{\pi}_1 + \tilde{\pi}_2}.
\]
\item With probability $\pi$, draw $\alpha \sim \text{Gamma}(a + k, b - \log z)$; otherwise draw
\[
\alpha \sim \text{Gamma}(a + k - 1, b - \log z).
\]
\end{itemize}

When fitting a DP, the value of $\alpha$ can be sampled by default in the \pkg{dirichletprocess} package.

\subsubsection{Inferring the Base Measure}

The base measure $G_0$ may itself depend on hyperparameters $\gamma$, in which case we write it as $G_0(\cdot \mid \gamma)$ and place a prior distribution $p(\gamma)$ on these quantities. This leads to the hierarchical specification
\begin{align*}
\phi_i \mid \gamma &\sim G_0, \\
\gamma &\sim p(\gamma),
\end{align*}
so that inference for $\gamma$ is based on its posterior distribution
\[
p(\gamma \mid \phi_{1:m}),
\]
where $\phi_{1:m}$ denotes the distinct cluster parameters in the current state of the sampler.

When this posterior is available in closed form, $\gamma$ can be updated directly. Otherwise, a Metropolis--Hastings step may be used. In the current implementation, such base-measure hyperparameter updates are provided only for selected mixture models rather than uniformly across all kernels.

\subsection{Implemented Mixture Models} \label{subsec:implemented}
As we will discuss, one of the strengths of the \pkg{dirichletprocess} package is that it allows users to specify DPMs using whichever choices of the kernel $k$ and base measure $G_0$ they please. However for ease of use, we have implemented certain choices of $k$ and $G_0$ directly in the
package, including routines for resampling hyperparameters where these updates are implemented.

\subsubsection{Gaussian Mixture Model}
The Gaussian distribution is the most commonly used mixture model. In this case, $\theta = (\mu,\sigma^2)$, where $\mu$ and $\sigma^2$ are the mean and variance respectively. The kernel is:
\begin{equation*}
k(y \mid \theta) = N(y_i \mid \mu, \sigma^2) = \frac{1}{\sqrt{2\pi \sigma^2}} \exp \left( \frac{ -(y-\mu)^2}{2\sigma^2} \right).
\end{equation*}
The conjugate prior for $\theta$ is the Normal-Inverse-Gamma distribution, with parameters $\gamma = (\mu_0, \kappa_0, \alpha_0, \beta_0)$:
\begin{equation*}
G_0(\theta \mid \gamma) =  N\left(\mu \mid \mu_0, \frac{\sigma^2}{\kappa_0} \right) \text{Inv-Gamma} \left(\sigma^2 \mid \alpha_0, \beta_0 \right).
\end{equation*}
The default parameter values are $\mu_0 = 0$, $\sigma_0^2 = 1$, $\alpha_0 = 1$, and $\beta_0 = 1$.
We recommend rescaling the data $y$ so that its mean is 0 and standard deviation is 1. This makes the default parameterisation of $G_0$ relatively uninformative.

Since this prior is conjugate, the predictive distribution for a new observation $\tilde{y}$ can be found analytically, and is a location-scale Student-$t$ distribution:
\begin{equation*}
p(\tilde{y} \mid \gamma) = \int k(\tilde{y}\mid\theta) p(\theta \mid G_0) d\theta = \frac{1}{\tilde{\sigma}} \text{Student-t} \left( \frac{ \tilde{y} - \tilde{\mu}}{\tilde{\sigma}} \mid \tilde{v}\right),
\end{equation*}
where $\tilde{v} = 2 \alpha_0$, $\tilde{\mu} = \mu_0$, and $\tilde{\sigma} = \sqrt{ \frac{\beta_0 (\kappa_0 + 1)}{\alpha_0 \kappa_0}}$.

Finally, the posterior distribution is also Normal-Inverse-Gamma due to the conjugacy of the prior:
\begin{align*}
p(\theta \mid y, \gamma) & = N \left( \mu \mid \mu_n , \frac{\sigma^2}{\kappa_0 + n}  \right) \text{Inv-Gamma} \left( \sigma^2 \mid \alpha_n , \beta_n \right), \\
\mu_n & = \frac{\kappa_0 \mu_0 + n \overline{y}}{\kappa_0 + n}, \\
\alpha_n & = \alpha_0 + \frac{n}{2}, \\
\beta_n & = \beta_0 + \frac{1}{2} \sum_{i=1}^n (y_i - \overline{y})^2 + \frac{\kappa_0 n (\overline{y} - \mu_0)^2}{2(\kappa_0 + n)} .
\end{align*}

\subsubsection{Multivariate Gaussian Mixture Model - Conjugate}

The multivariate Gaussian mixture model is one of the most widely used nonparametric modelling
approaches for multivariate data, and is also heavily used in clustering applications
\citep{maceachern_estimating_1998}. For $d$-dimensional data, the unknown parameter is
$\theta = (\mu, \Sigma)$, where $\mu$ is a column vector of length $d$ and $\Sigma$ is a
$d \times d$ covariance matrix. The kernel is:
\begin{equation*}
k(y_i \mid \theta) = \frac{1}{(2\pi)^{d/2} |\Sigma|^{1/2}}
\exp \left( -\frac{1}{2} (y_i - \mu)^\top \Sigma^{-1} (y_i - \mu) \right).
\end{equation*}

For the prior choice we use a Normal-Inverse-Wishart distribution:
\begin{equation*}
G_0(\mu, \Sigma \mid \mu_0, \kappa_0, \nu_0, \Phi_0)
=
N \left( \mu \mid \mu_0, \frac{\Sigma}{\kappa_0} \right)
IW_{\nu_0}(\Sigma \mid \Phi_0),
\end{equation*}
where $\mu_0$ is the prior mean vector, $\kappa_0$ and $\nu_0$ are scalar hyperparameters,
and $\Phi_0$ is a positive definite scale matrix. The default prior parameter values are
$\mu_0 = 0$, $\Phi_0 = I$, $\kappa_0 = d$, and $\nu_0 = d$.

This prior is conjugate, so the posterior distribution can be written analytically:
\begin{align*}
p(\theta \mid y)
&=
N \left( \mu \mid \mu_n, \frac{\Sigma}{\kappa_n} \right)
IW_{\nu_n}(\Sigma \mid \Phi_n), \\
\mu_n
&=
\frac{\kappa_0 \mu_0 + n \bar{y}}{\kappa_0 + n}, \\
\kappa_n
&=
\kappa_0 + n, \\
\nu_n
&=
\nu_0 + n, \\
\Phi_n
&=
\Phi_0 +
\sum_{i=1}^n (y_i - \bar{y})(y_i - \bar{y})^\top +
\frac{\kappa_0 n}{\kappa_0 + n}(\bar{y} - \mu_0)(\bar{y} - \mu_0)^\top.
\end{align*}

Again, as this is a conjugate mixture, the predictive distribution for a new observation
$\tilde{y}$ can be found analytically and is a multivariate Student-$t$ distribution.

\subsubsection{Multivariate Gaussian Mixture Model -- Semi-Conjugate}
In the semi-conjugate case, the base measure for $(\mu, \Sigma)$ is specified independently as
\begin{equation*}
G_0(\mu, \Sigma) = N(\mu \mid \mu_0, \Sigma_0)\, IW_{\nu_0}(\Sigma \mid \Phi_0).
\end{equation*}
Since this prior is not jointly conjugate, sampling from the posterior is carried out using the
conditional posterior distributions. These are
\begin{align*}
\Sigma \mid \mu, \nu_0, \Phi_0
&\sim IW_{\nu_n}(\Phi_n), \\
\nu_n
&= \nu_0 + n, \\
\Phi_n
&= \Phi_0 + \sum_{i=1}^n (y_i - \mu)(y_i - \mu)^\top,
\end{align*}
and
\begin{align*}
\mu \mid \Sigma, \mu_0, \Sigma_0
&\sim N(\mu_n, \Sigma_n), \\
\Sigma_n
&= \left(\Sigma_0^{-1} + n \Sigma^{-1}\right)^{-1}, \\
\mu_n
&= \Sigma_n \left(\Sigma_0^{-1}\mu_0 + n \Sigma^{-1}\bar{y}\right).
\end{align*}
Using these conditional distributions, each parameter can be sampled using the previous
sample. This allows us to use Algorithm 8 and treat the model as a non-conjugate mixture model.

\subsubsection{Beta Mixture Model}
Dirichlet process mixtures of Beta distributions have been considered by \citet{kottas_nonparametric_2006}
for the nonparametric estimation of continuous distributions defined on a bounded interval $[0, T]$.
For ease of interpretation, we follow their parameterisation of the Beta distribution in terms of its
mean $\mu$ and precision parameter $\nu$. In this case, $\theta = (\mu, \nu)$ with known parameter $T$.
The mixture kernel is:
\begin{equation*}
k(y_i \mid \theta) = \text{Beta}(y_i \mid \mu, \nu, T) =
\frac{y_i^{\mu \nu / T - 1}(T-y_i)^{\nu(1-\mu/T)-1}}
{B(\mu \nu / T, \nu(1-\mu/T))T^{\nu-1}}.
\end{equation*}

There is no conjugate prior for this mixture kernel. Instead, the \pkg{dirichletprocess} package
uses the non-conjugate prior from \citet{kottas_nonparametric_2006}, where
\begin{equation*}
G_0(\mu, \nu \mid T, \alpha_0, \beta_0) =
U(\mu \mid [0,T]) \, \text{Inv-Gamma}(\nu \mid \alpha_0, \beta_0).
\end{equation*}

To sample from the posterior distribution we use the Metropolis-Hastings algorithm, with
default parameters $\alpha_0 = 2$ and $\beta_0 = 8$. There is also the option to place a prior
distribution on $\beta_0$ and update this hyperparameter with each iteration. For this we use the
default prior
\begin{equation*}
\beta_0 \sim \text{Gamma}(a,b),
\end{equation*}
with $a = 1$ and $b = 0.125$ by default.

\subsubsection{Weibull Mixture Model}
The Weibull distribution has strictly positive support and is mainly used for modelling
positive-valued data. It is also widely used in survival analysis. Mixtures of Weibull
distributions have been considered by \citet{kottas_nonparametric_2006} for a variety of
survival applications. The parameters of the Weibull distribution are the shape $a$ and scale $b$:
\begin{equation*}
k(y_i \mid \theta) = \text{Weibull}(y_i \mid a, b) = \frac{a}{b} y_i^{a-1}
\exp \left( -\frac{y_i^a}{b} \right),
\end{equation*}
where $\theta = (a,b)$.

We use the non-conjugate Uniform--Inverse-Gamma prior
\begin{equation*}
G_0(a,b \mid \phi,\alpha,\beta) = U(a \mid 0,\phi)\,\text{Inv-Gamma}(b \mid \alpha,\beta),
\end{equation*}
for the unknown parameters. By default, $\phi$, $\alpha$, and $\beta$ are not all treated in the
same way: $\alpha$ is fixed, while priors are placed on $\phi$ and $\beta$ and these are updated
during fitting.

For $\phi$ we use a Pareto prior, which is conjugate to the Uniform distribution:
\begin{align*}
a_i \mid \phi &\sim U(0,\phi), \\
\phi &\sim \text{Pareto}(x_m, k), \\
\phi \mid a_i &\sim \text{Pareto}(\max\{a_i,x_m\}, k+n).
\end{align*}
By default $x_m = 6$ and $k = 2$, giving a prior with infinite variance.

As $b$ has an Inverse-Gamma distribution with fixed shape $\alpha$, a conjugate Gamma prior
can be placed on $\beta$:
\begin{align*}
b_i \mid \beta &\sim \text{Inv-Gamma}(\alpha,\beta), \\
\beta &\sim \text{Gamma}(\alpha_0,\beta_0), \\
\beta \mid b &\sim \text{Gamma}\left(\alpha_0 + n\alpha,\ \beta_0 + \sum_{i=1}^n \frac{1}{b_i}\right).
\end{align*}
Here $\alpha$ is fixed by the user, while $\alpha_0 = 1$ and $\beta_0 = 0.5$ by default.
As this is a non-conjugate mixture model, a Metropolis-Hastings step is used to sample from
the posterior distribution.

\section{Package Overview}
\label{sec:package}

The \pkg{dirichletprocess} package contains implementations of a variety of Dirichlet process mixture models for nonparametric Bayesian analysis. Unlike several other \proglang{R} packages, the emphasis is less on providing a set of functions which completely automate routine tasks (e.g. density estimation or linear regression) and more on providing an abstract data type representation of Dirichlet process objects which allow them to be used as building blocks within hierarchical models.

To illustrate how the package is meant to be used, and how it differs from other \proglang{R} packages, consider the task of density estimation. Suppose we wish to estimate the density of some data stored in the variable \code{y} using a Dirichlet process mixture of Gaussian distributions. This is done as follows:

\begin{CodeInput}
R> y <- rt(200, 3) + 2 #generate sample data
R> dp <- DirichletProcessGaussian(y)
R> dp <- Fit(dp, 1000)
\end{CodeInput}
The function \code{DirichletProcessGaussian} is the creator function for a mixture model of univariate Gaussians. This creates the object \code{dp}. We then use \code{Fit} to infer the cluster allocations and parameters using the Chinese Restaurant Process sampler described in Section~\ref{subsec:dpm}. At each iteration the cluster labels are updated, the cluster parameters are resampled, and the concentration parameter $\alpha$ is updated. Thus, for standard mixture models, \code{Fit} provides an out-of-the-box way to fit a Dirichlet process mixture model after specifying only the mixture kernel.

By default, \code{Fit} retains the generated MCMC samples inside the fitted object. This behaviour is controlled by the arguments \code{storeSamples} and \code{thinning}. The default \code{storeSamples = TRUE} retains samples in memory, while \code{storeSamples = FALSE} runs the sampler and updates the current fitted state without appending any new sample history. The argument \code{thinning = k} retains iterations \code{1, 1+k, 1+2k, \ldots} from the current call to \code{Fit}. If \code{Fit} is called repeatedly, newly retained samples are appended to the existing retained history; previous samples are not re-thinned or deleted.

The retained samples can be summarised using \code{PosteriorSummary}, which computes posterior means, medians, and pointwise credible intervals. When usable retained samples are available, \code{plot(dp)} uses these summaries by default; otherwise it plots only the current fitted curve.

The \pkg{dirichletprocess} package currently provides the following features:

\begin{itemize}
\item Implementations of Dirichlet process mixture models using Gaussian (both univariate and multivariate), Beta, and Weibull mixture kernels.
\item Implementation of various schemes for re-sampling model parameters.
\item Access to samples from the Dirichlet process in both marginal form, as well as in (truncated) stick-breaking form
\item A flexible way for the user to add new Dirichlet process models which are not currently implemented in the package, and yet still use the re-sampling functions from the package. To illustrate this, Section \ref{sec:NewMDobj} shows how simple it is to create a Dirichlet process mixture model with Poisson and Gamma distribution kernels, even though this is not implemented in the package.
\item An ability to plot fitted densities, posterior summaries, and pointwise credible intervals using \code{plot}.
\end{itemize}

All of the above features will be demonstrated in the following examples. Before running the examples below, we load the \pkg{dirichletprocess} package together with the plotting and simulation packages used in the examples.

\begin{CodeInput}
R> library("dirichletprocess")
R> library("ggplot2")
R> library("mvtnorm")
\end{CodeInput}

\subsubsection{Nonparametric Density Estimation}
The simplest application of DPMMs is to  nonparametrically estimate the distribution of independent and identically distributed observations $y_1,\ldots,y_n$, where:
\begin{align*}
y_i & \sim F ,\\
F & = \sum _{i=1} ^n \pi _i k(y_i \mid \theta _i),
\end{align*}
where $k$ is some density function parameterised by $\theta _i$ and $n$ is an unknown number of clusters (i.e. F has been specified nonparametrically). The most widely used specification is the Gaussian mixture kernel with a Normal-Inverse Gamma base measure $G_0$, which is described more fully in Section \ref{subsec:implemented}.

As an example we use the waiting times between eruptions of the Old Faithful volcano. This dataset is available within \proglang{R} and called \code{faithful}. We transform the waiting times to be zero mean  and unit standard deviation and proceed to fit a DPMM with the default settings. We then model the waiting times as a mixture of Normal distributions

\begin{align*}
y_i & \sim F, \\
F & = \sum _{i=1} ^n \pi _i k(y_i \mid \theta _i) \quad \theta _i = \left\lbrace \mu _i , \sigma ^2 _i \right\rbrace , \\
\theta _i & \sim G, \\
G & \sim \text{DP}(\alpha , G_0),
\end{align*}
where $k(y _i \mid \theta )$ is the standard normal probability density and $G_0$ is the base measure as in Section \ref{subsec:implemented}.

\begin{CodeInput}
R> its <- 500
R> faithfulTransformed <- scale(faithful$waiting)
R> dp <- DirichletProcessGaussian(faithfulTransformed)
R> dp <- Fit(dp, its)
R> plot(dp)
R> plot(dp, data_method="hist")
\end{CodeInput}
\begin{figure}[tb]
\centering
  \begin{subfigure}[t]{0.45\textwidth}
	  \includegraphics[width=0.95\textwidth]{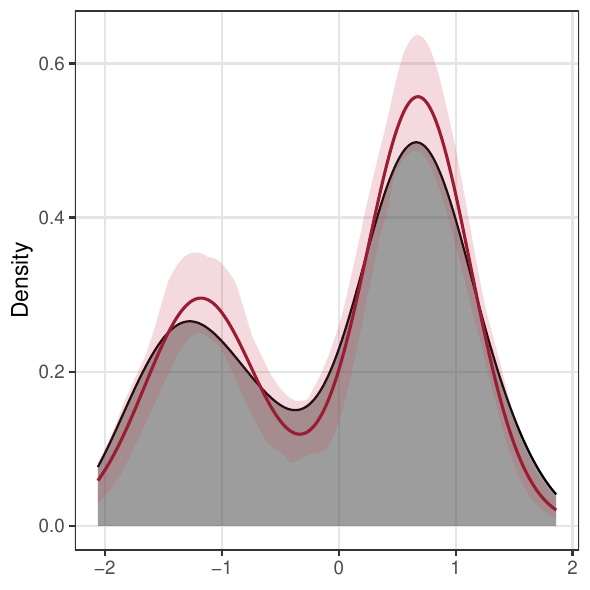}
	  \caption{The estimated density of the data is plotted with the DPMM posterior mean and pointwise credible intervals from retained MCMC samples overlaid in red. }
	  \label{fig:oldfaithfuldens}
	\end{subfigure} \hfill
	\begin{subfigure}[t]{0.45\textwidth}
	  \includegraphics[width=0.95\textwidth]{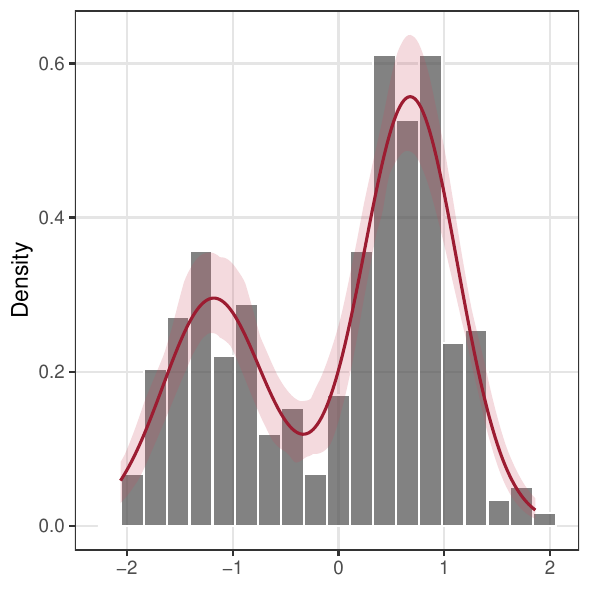}
	  \caption{Instead of a density estimate, a histogram is plotted for the data.}
	  \label{fig:oldfaithfulhist}
	\end{subfigure}
  \caption{Old Faithful waiting times density estimation with a DPMM of Gaussians.}
  \label{fig:oldfaithful}
\end{figure}
The resulting fit from the model is shown in Figure \ref{fig:oldfaithful}. The multi-modal nature has been successfully captured.

When plotting the resulting object, the \code{plot} function has the argument \code{data_method} which controls how the data is represented in the plot. In the default case it is a density estimation. For cases where the amount of data is limited, a density estimation may not be suitable, in which case, \code{data_method} should be set to \code{"hist"} for a histogram.
% After running the \code{Fit} function, the DP object contains 3 MCMC parameter chains by default:
% \begin{itemize}
% \item A list containing the samples of the cluster parameters $\phi _i$.
% \item A list containing the weights associated with each cluster parameter $w_i$.
% \item A numeric vector of the samples of $\alpha$.
% \end{itemize}
% Using these quantities, the user can then perform the appropriate analysis.

For most users the \code{Fit} function is sufficient for practical purposes. However, for the more advanced users who wish to alter how they fit the dirichletprocess object there are a number of functions available to help.

By default, the \code{Fit} function updates the cluster allocation, the cluster parameters and then the $\alpha$ parameter. In some rare cases, updating $\alpha$ every iteration can delay convergence. Instead, the user can instead choose to update $\alpha$ every 10 iterations.
\newpage %stop splitting
\begin{CodeInput}
R> dp <- DirichletProcessGaussian(y)
R> 
R> samples <- list()
R> for(s in seq_len(1000)){
+   dp <- ClusterComponentUpdate(dp)
+   dp <- ClusterParameterUpdate(dp)
+ 
+   if(s %% 10 == 0) {
+   	dp <- UpdateAlpha(dp)
+   }
+   samples[[s]] <- list()
+   samples[[s]]$phi <- dp$clusterParameters
+   samples[[s]]$weights <- dp$weights
+ }
\end{CodeInput}
The function \code{ClusterComponentUpdate} iterates through all the data points, $y_i$ for $i=1,\ldots,n$, updating its cluster assignment sequentially via the Chinese Restaurant Process sampling algorithm, using Key Property 4 in Section \ref{subsec:dpm}. For each data point, it can either be assigned to an existing cluster, or form a new cluster. The probability it is assigned to an existing cluster is proportional to $n_i k(y _j \mid \theta _{i})$, where $n_i$ is the number of points already assigned to the cluster $\theta _i$ and $k$ is the likelihood of the data point  evaluated with the cluster parameter $\theta _i$. The probability that it forms a new cluster is proportional to $\alpha$, the concentration parameter. If the datapoint is selected to form a new cluster, then the corresponding new cluster
parameter $\theta_{\text{new}}$ is drawn from the appropriate conditional distribution:
in the conjugate case this is the posterior distribution implied by Key Property 4, while
in the non-conjugate case the sampling is carried out using Algorithm 8 from \citet{neal_markov_2000}.
Subsequent points can now also be added to this cluster. Note that when a conjugate base measure is used in the DP, this function samples directly from the conditional posteriors, while if a non-conjugate sampler is used then the sampling is carried out using Algorithm 8 from \cite{neal_markov_2000}.

After each data point has its cluster allocation updated the function \code{ClusterParameterUpdate} is called and resamples each of the unique $\theta _j$ parameters (i.e. if there are $m$ unique values/clusters in $\theta_1,\ldots,\theta_m$ then all $m$ are resampled). The new values of $\theta_j$ are sampled from the posterior distribution of the parameter using all the data associated to that cluster parameter as per Equation \eqref{eq:clusterposterior}.

Finally, \code{UpdateAlpha} samples a new value of $\alpha$ from its posterior distribution using the method outlined in Section \ref{subsec:hyperparam}. By manually calling these functions the user has control over the MCMC routine without having to have specific knowledge of the required algorithms.

The key point of the  \pkg{dirichletprocess} package which the above code highlights is that a) the user controls when to re-sample the DP parameters, and b) the current sample is contained in the DP object and ready for inspection at any point in the code. This allows DP objects to be used as building blocks within hierarchical models. Unlike calls to \code{Fit}, however, manually calling the update functions does not automatically append states to the retained sample history. In the example above, the list \code{samples} is created explicitly to store the quantities of interest from each manual iteration.

For efficiency, the core DP sampling algorithms are implemented in \proglang{C++}. As such, the \code{Fit} function can carry out the main sampling steps efficiently. When the same updates are instead called manually inside an \proglang{R} loop (as above), performance is typically somewhat slower due to the repeated overhead of transferring control between the \proglang{R} wrapper and the compiled \proglang{C++} code.

\subsection{Density Estimation on Bounded Intervals}
In some situations it will be necessary to estimate densities on bounded intervals. For example, it might be known that the observations $y_i$ are restricted to lie within the interval $[0,1]$. In this case, a mixture of Gaussian distributions is inappropriate, since this will assign positive probability to the whole real line. An alternative specification is a mixture of Beta distributions, since the Beta distribution only has positive mass in $[0,1]$. The full model is the same as in the previous example but replacing $k$ with the Beta distribution.

\begin{CodeInput}
R> y <- c(rbeta(150, 1, 3), rbeta(150, 7, 3)) #generate sample data
R> dp <- DirichletProcessBeta(y, 1)
R> dp <- Fit(dp, 1000)
\end{CodeInput}

As the data in this example are simulated, we can compare the fitted density to the true generating density. We evaluate \code{PosteriorSummary} on a grid, which uses the retained MCMC samples stored by \code{Fit} to compute posterior means and pointwise credible intervals. The arguments \code{burnin} and \code{thinning} below are applied only when constructing this summary; they do not alter the fitted object itself.

\begin{CodeInput}
R> xGrid <- ppoints(100)
R> posteriorSummary <- PosteriorSummary(dp, xGrid,
+                                       burnin = 500,
+                                       thinning = 5,
+                                       level = 0.95)
R> 
R> trueFrame <- data.frame(x=xGrid,
+                         y=0.5*dbeta(xGrid, 1, 3)+
+                           0.5*dbeta(xGrid, 7, 3))
R> 
R> ggplot() +
+   geom_ribbon(data=posteriorSummary,
+               aes(x=x, ymin=Lower, ymax=Upper),
+               alpha=0.2,
+               colour=NA,
+               fill="red") +
+   geom_line(data=posteriorSummary, aes(x=x, y=Mean), colour="red") +
+   geom_line(data=trueFrame, aes(x=x, y=y))
\end{CodeInput}
\begin{figure}[tb]
	\centering
	\includegraphics[width=0.45\textwidth]{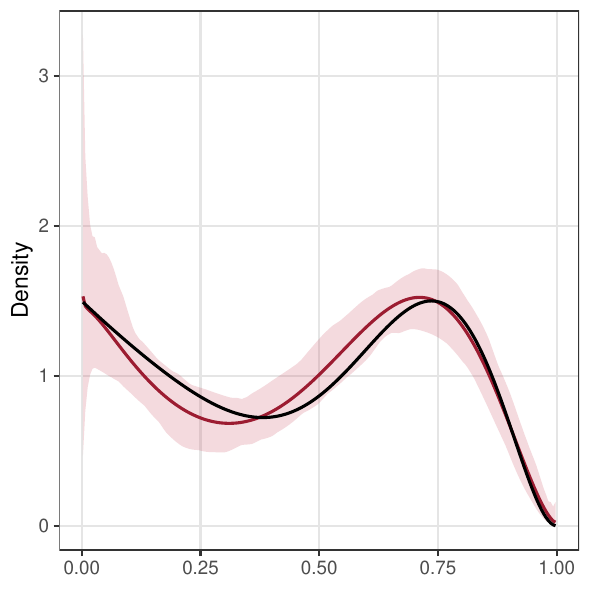}
	\caption{Posterior mean and pointwise credible intervals from retained MCMC samples using a Beta Dirichlet process mixture model.}
	\label{fig:densitybounded}
\end{figure}
Figure \ref{fig:densitybounded} shows the resulting posterior distribution and true density.

\subsection{Cluster Analysis (Multivariate)} \label{subsec:clusteranalysis}
For any Dirichlet model, each data point $y_i$ is assigned a cluster parameter $\theta _i$. The collection of unique values of cluster parameters $\theta _i ^*$ allows for a natural way of grouping the data and hence the Dirichlet process is an effective way of performing cluster analysis. For multidimensional data it is most common to use a mixture of multivariate normal distributions to cluster the observations into appropriate groups. In the \pkg{dirichletprocess} package, the clustering labels are available to the user during each fitting iteration as \code{dp$clusterLabels}. Examples of the use of Dirichlet processes in clustering can be found in \cite{teh_sharing_2005} and \cite{kim_variable_2006}.

To demonstrate this we return to the \code{faithful} dataset which consists of two-dimensional data. In this example we also consider the length of the eruption as well as the amount of time between eruptions. The full model can be written as
\begin{align*}
y _i & \sim N (y \mid \theta_i ), \\
\theta_i & =  \left\lbrace \boldsymbol{\mu}_i, \Sigma _i \right\rbrace , \\
\theta _i & \sim G, \\
G & \sim \text{DP} ( \alpha, G_0 ),
\end{align*}
where the prior parameters of $G_0$ take on their default value as shown in Section \ref{subsec:implemented}. We will be using the cluster labels to indicate which group each data point belongs to.

We transform the data so that each variable is zero mean and unit standard deviation.
\begin{CodeInput}
R> faithfulTrans <- scale(faithful)
\end{CodeInput}
We form the dirichletprocess object and perform 1000 MCMC samples.
\begin{CodeInput}
R> dp <-  DirichletProcessMvnormal(faithfulTrans)
R> dp <- Fit(dp, 1000)
R> plot(dp)
\end{CodeInput}
When analysing the results of the fit we are interested in the final Gibbs sample of the cluster labels. Using the cluster label to assign a colour to each datapoint we can easily see the distinct clusters that form.

\begin{figure}[tb]
  \centering
	\includegraphics[width=0.5\textwidth]{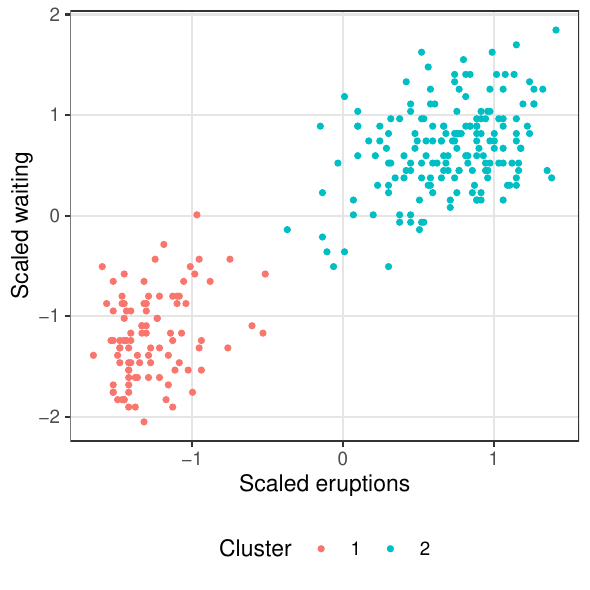}
	\caption{The colour of the points indicates that there are groups in the \code{faithful} dataset.}
	\label{fig:clustering}
\end{figure}

Here Figure \ref{fig:clustering} shows the last iteration of the cluster labels and the colours indicate the found clusters. Whilst this example only shows two dimensions, the code is generalised to work with as many dimensions as necessary.

\subsection{Modifying the Observations}
In some applications of Dirichlet processes the data being used can change from iteration to iteration of the sampling algorithm. This could be because the values of the data change, or because for a full data set $\mathbf{y} = y_1 , \ldots , y_n$, only subsets of the data are used at each iteration. When fitting a DP object we provide the appropriate function \code{ChangeObservations} to change the observations between iterations.

This function takes the new data, predicts what clusters from the previous fitting the new data belongs to and updates the clustering labels and parameters accordingly. A modified object with the new data associated to clusters and the function \code{Fit} is ready to be used to sample the cluster parameters and weights again.

\subsubsection{Example: Priors in Hierarchical Models}
One application of observations changing with each iteration is using a Dirichlet process as a prior for a parameter in a hierarchical model. An example of hierarchical modelling comes from \cite{gelman_bayesian_2014} involving tumour risk in rats. In this example, there are 71 different experiments, and during each experiment a number of rats are inspected for tumours, with the number of tumours in each experiment being the observed data. This data is included in our package and can be found in the \textbf{rats} variable, with the first column being the number of tumours in each experiment, and the second being the number of rats.

A naive approach would model each experiment as a Binomial draw with unknown $\theta _i$ and known $N_i$. A Beta distribution is the conjugate prior for the Binomial distribution and would be used as the prior on $\theta$:
\begin{align*}
y_i \mid \theta_i, N_i \sim & \text{Binomial}(N_i,\theta_i), \\
\theta_i \sim & \text{Beta}(\alpha, \beta).
\end{align*}

\begin{figure}[tb]
\begin{subfigure}[b]{0.5\textwidth}
	\centering
	\includegraphics[width=0.9\textwidth]{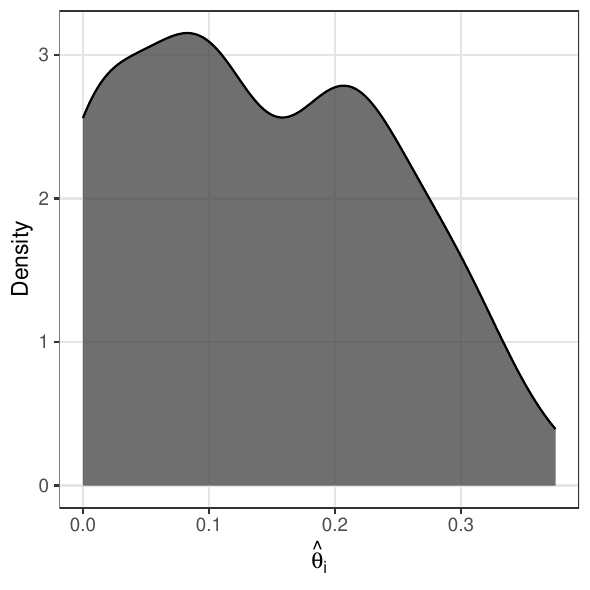}
	\caption{Empirical distribution of $\hat{\theta}_i$.}
	\label{fig:ratsImperical}
\end{subfigure}
\begin{subfigure}[b]{0.5\textwidth}
	\centering
	\includegraphics[width=0.9\textwidth]{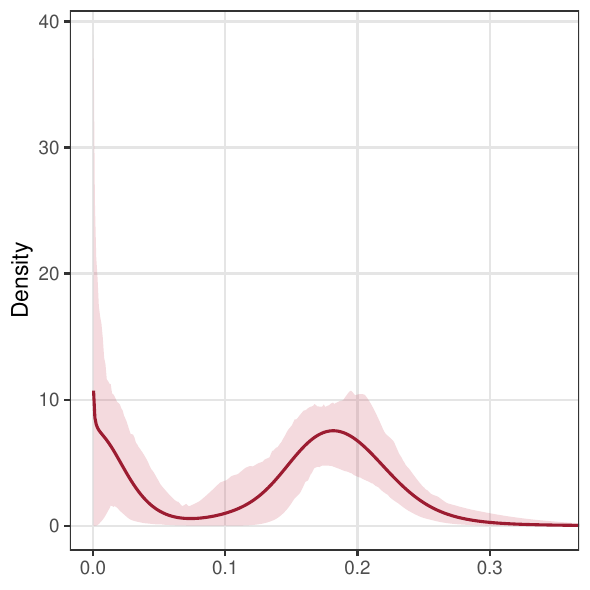}
	\caption{Dirichlet prior model for $p(\theta _i)$.}
	\label{fig:rats_dirichlet_model}
\end{subfigure}
\caption{Rat tumour risk empirical density and Dirichlet process prior estimate with pointwise credible intervals from retained MCMC samples.}
\end{figure}

However, Figure \ref{fig:ratsImperical} shows the empirical distribution of $\hat{\theta}_i = \frac{y_i}{N_i}$. This distribution shows hints of bimodality, something that a single Beta distribution cannot capture and hence the prior choice of $p(\theta_i)$ is dubious.
An alternative procedure is to instead use a nonparametric prior on the $\theta_i's$. Since these parameters are constrained to lie between 0 and 1, one choice might be a Dirichlet process mixture of Beta distributions. As discussed in Section~2.3, the Beta mixture model in the \pkg{dirichletprocess} package is parameterised in terms of its mean $\mu_i$ and precision $\tau_i$, rather than the usual pair of shape parameters. This leads to the following model

\begin{align*}
y_i \mid \theta_i, N_i \sim & \text{Binomial}(N_i,\theta_i), \\
\theta_i \sim & \text{Beta}(\mu_i, \tau_i), \\
\mu _i, \tau _i \sim & F, \\
F \sim & DP(\alpha, G_0),
\end{align*}
where $\alpha$ and $G_0$ follow the default implementations of the \pkg{dirichletprocess} package. We then implement this model using the \pkg{dirichletprocess} functions as follows:
\clearpage

\begin{CodeInput}
R> numSamples = 200
R> thetaDirichlet <- matrix(nrow=numSamples, ncol=nrow(rats))
R> 
R> dpobj <- DirichletProcessBeta(rats$y/rats$N,
+                               maxY=1,
+                               g0Priors = c(2, 150),
+                               mhStep=c(0.25, 0.25),
+                               hyperPriorParameters = c(1, 1/150))
R> dpobj <- Fit(dpobj, 10)
R> 
R> clusters <- dpobj$clusterParameters
R> 
R> shape1 <- clusters[[1]] * clusters[[2]]
R> shape2 <- (1 - clusters[[1]]) * clusters[[2]]
R> 
R> for(i in seq_len(numSamples)){
+ 
+   posteriorShape1 <- shape1[dpobj$clusterLabels] + rats$y
+   posteriorShape2 <- shape2[dpobj$clusterLabels] + rats$N - rats$y
+   thetaDirichlet[i, ] <- rbeta(nrow(rats), posteriorShape1, posteriorShape2)
+ 
+   dpobj <- ChangeObservations(dpobj, thetaDirichlet[i, ])
+   dpobj <- Fit(dpobj, 5, progressBar=FALSE)
+   clusters <- dpobj$clusterParameters
+ 
+   shape1 <- clusters[[1]] * clusters[[2]]
+   shape2 <- (1 - clusters[[1]]) * clusters[[2]]
+ }
\end{CodeInput}
Note the reason why the observations are changing is because the DP mixture model is applied to the $\theta_i$ parameters, which are resampled (and hence have different values) during each MCMC iteration.
Because \code{Fit} retains samples by default, each call to \code{Fit(dpobj, 5, progressBar = FALSE)} appends its newly retained iterations to the existing retained history. Earlier retained samples are not re-thinned or deleted. In this example the observations are themselves updated inside the loop, so the summary below is based on the retained states produced during the full iterative scheme rather than from a single fixed-data call to \code{Fit}.

\begin{CodeInput}
R> ggplot(rats, aes(x=y/N)) +
+   geom_density(fill="black") #Plot the empirical distribution
R> 
R> 
R> posteriorSummary <- PosteriorSummary(dpobj, ppoints(1000),
+                                      burnin = 500,
+                                      thinning = 5,
+                                      level = 0.90)
R> 
R>   ggplot() +
+          geom_ribbon(data=posteriorSummary,
+                      aes(x=x, ymin=Lower, ymax=Upper),
+                      alpha=0.2) +
+          geom_line(data=posteriorSummary, aes(x=x, y=Mean)) +
+          coord_cartesian(xlim = c(0, 0.35))

\end{CodeInput}
Plotting the resulting estimation in Figure \ref{fig:rats_dirichlet_model} reveals that the DP is a more suitable prior than the Beta distribution. This confirms what we saw from the empirical distribution that the data is bi-modal.

\subsection{Hierarchical Dirichlet process}

A hierarchical Dirichlet process (HDP) \citep{teh_sharing_2005} can be used for grouped data,
where each dataset is modelled using its own Dirichlet process, but the group-specific
Dirichlet processes are linked through a shared global base measure. This allows mixture
components to be shared across groups while still permitting each group to have its own
mixing proportions. Mathematically, the model can be written as
\[
y_{ij} \sim F(\theta_{ij}),
\]
\[
\theta_{ij} \sim G_j,
\]
\[
G_j \sim DP(\alpha_j, G_0),
\]
\[
G_0 \sim DP(\gamma, H),
\]
for each dataset $j = 1, \ldots, J$.

The key feature of this model is that the group-specific measures $G_j$ share atoms through
the common global measure $G_0$. This allows the same mixture component to appear in
multiple datasets, while still allowing the relative importance of that component to vary from
group to group.

In the \texttt{dirichletprocess} package, posterior inference for the hierarchical model is
based on an explicit franchise-style representation. Within each group, observations are
allocated to local tables, and each occupied table is assigned to a shared global dish. The
global dish parameters are then updated using the pooled data assigned to that dish across all
groups. This provides a natural way to represent the sharing of mixture components between
datasets, while retaining group-specific mixture weights through the local table allocations.
The local concentration parameters $\alpha_j$ and the global concentration parameter
$\gamma$ are updated from the corresponding occupied-table and active-dish counts.

In this example we create two synthetic data sets and fit a hierarchical Dirichlet process to
demonstrate the use of such a model. In this case we are fitting a Beta Dirichlet mixture model.
As in Section~2.3, the Beta components are parameterised by their mean $\mu$ and precision
$\tau$. We therefore simulate from two Beta mixtures
\[
y_1 \sim 0.5 \text{Beta}(\mu=0.25, \tau=5) + 0.5 \text{Beta}(\mu=0.75, \tau=6),
\]
\[
y_2 \sim 0.5 \text{Beta}(\mu=0.25, \tau=5) + 0.5 \text{Beta}(\mu=0.4, \tau=10),
\]
where there is a common group of parameters between the two datasets. First we simulate
the two data sets.

\begin{CodeInput}
R> mu <- c(0.25, 0.75, 0.4)
R> tau <- c(5, 6, 10)
R> shape1 <- mu * tau
R> shape2 <- (1 - mu) * tau
R> y1 <- c(rbeta(500, shape1[1], shape2[1]), rbeta(500, shape1[2], shape2[2]))
R> y2 <- c(rbeta(500, shape1[1], shape2[1]), rbeta(500, shape1[3], shape2[3]))
\end{CodeInput}

We then use the appropriate constructor function to build a hierarchical Dirichlet object
with relatively uninformative priors for the global base distribution and fit for 2,000 iterations.

\begin{CodeInput}
R> dplist <- DirichletProcessHierarchicalBeta(list(y1, y2),
+                                            maxY=1,
+                                            hyperPriorParameters = c(1, 0.01),
+                                            mhStepSize = c(0.1, 0.1),
+                                            gammaPriors = c(2, 4),
+                                            alphaPriors = c(2, 4))
R> dplist <- Fit(dplist, 2000)
\end{CodeInput}

The constructor function \texttt{DirichletProcessHierarchicalBeta} returns a hierarchical
object containing one local \texttt{dirichletprocess} object for each dataset, together with the
shared global parameters and the top-level concentration parameter $\gamma$. The
\texttt{Fit} function updates the local allocations within each dataset, the shared global
components across datasets, and the concentration parameters governing the local and global
Dirichlet processes.

To visualise the fitted model we summarise the retained MCMC samples separately for each
group using the corresponding local Dirichlet process objects and compare these to the true
generating densities.

\newpage
\begin{CodeInput}
R> xGrid <- ppoints(100)
R> postSummary1 <- PosteriorSummary(dplist$indDP[[1]], xGrid,
+                                burnin = 1000,
+                                thinning = 5,
+                                level = 0.95)
R> postSummary1$Group <- "Group 1"
R> postSummary2 <- PosteriorSummary(dplist$indDP[[2]], xGrid,
+                                burnin = 1000,
+                                thinning = 5,
+                                level = 0.95)
R> postSummary2$Group <- "Group 2"
R> postSummary <- rbind(postSummary1, postSummary2)
R> 
R> trueFrame1 <- data.frame(y=0.5*dbeta(xGrid, shape1[1], shape2[1]) +
+                            0.5*dbeta(xGrid, shape1[2], shape2[2]),
+                          x=xGrid, Group="Group 1")
R> trueFrame2 <- data.frame(y=0.5*dbeta(xGrid, shape1[1], shape2[1]) +
+                            0.5*dbeta(xGrid, shape1[3], shape2[3]),
+                          x=xGrid, Group="Group 2")
R> trueFrame <- rbind(trueFrame1, trueFrame2)
R> 
R> ggplot() +
+   geom_ribbon(data=postSummary, aes(x=x, ymin=Lower, ymax=Upper),
+               alpha=0.2, colour=NA, fill="red") + #pointwise credible intervals
+   geom_line(data=postSummary, aes(x=x, y=Mean), colour="red") + #mean
+   geom_line(data=trueFrame, aes(x=x, y=y)) + #true density
+   facet_wrap(~Group)
\end{CodeInput}

\begin{figure}
\centering
\includegraphics[width=0.85\textwidth]{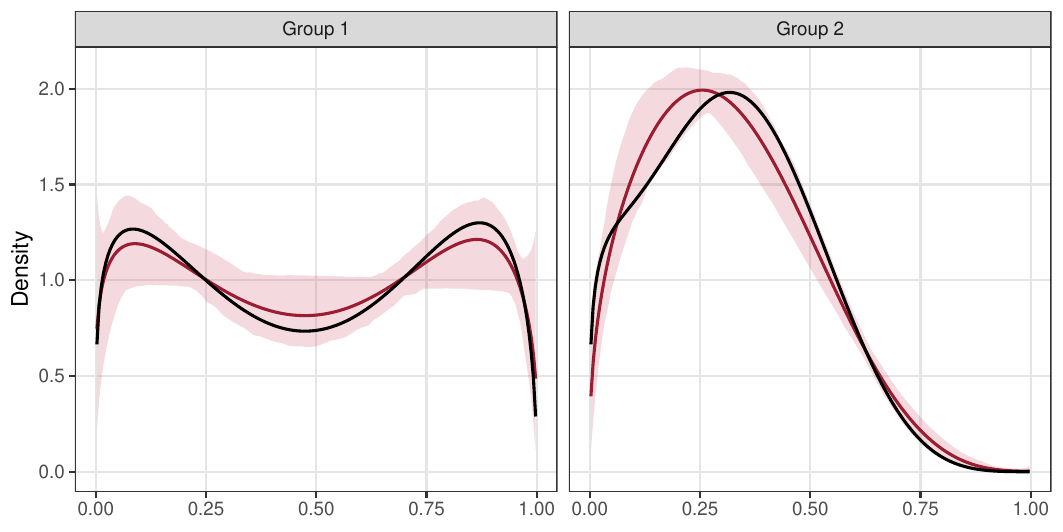}
\caption{Hierarchical Beta Dirichlet process mixture results. The black lines indicate the true generating distributions. The red lines and shaded areas are group-specific posterior means and pointwise credible intervals computed from the corresponding local Dirichlet process objects.}
\label{fig:hierBeta}
\end{figure}

The resulting group-specific posterior densities are plotted in Figure~\ref{fig:hierBeta}.
The fitted group-specific densities reflect the shared component between the two datasets while allowing each group to retain its own additional component.

\subsubsection{Hierarchical Multivariate Normal Distribution}

If the hierarchical data is not distributed across a bounded distribution or there are more than one dimension then the multivariate normal distribution might be more suited.

\begin{figure}
\centering
\includegraphics[width=0.85\textwidth]{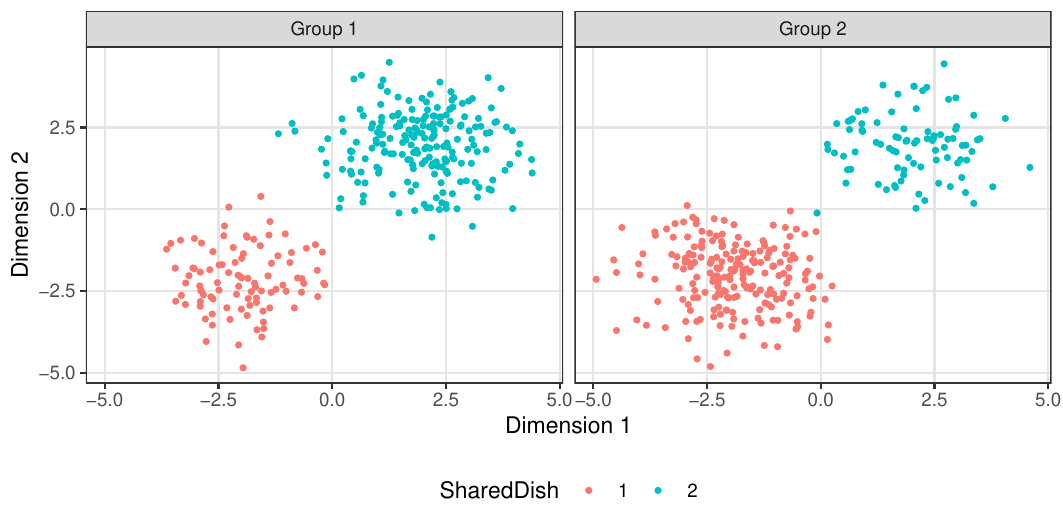}
\caption{Hierarchical multivariate normal Dirichlet results where the correct common clusters between the two datasets have been found.}
\label{fig:hierNormal}
\end{figure}

\begin{CodeInput}
R> N <- 300
R> 
R> #Sample N random uniform U
R> U <- runif(N)
R> 
R> group1 <- matrix(nrow=N, ncol=2)
R> group2 <- matrix(nrow=N, ncol=2)
R> #Sampling from the mixture
R> for(i in 1:N){
+   if(U[i]<.3){
+     group1[i,] <- rmvnorm(1,c(-2,-2))
+     group2[i,] <- rmvnorm(1,c(-2,-2))
+   }else if(U[i]<0.7){
+     group1[i,] <- rmvnorm(1,c(2,2))
+     group2[i,] <- rmvnorm(1,c(-2,-2))
+   }else {
+     group1[i,] <- rmvnorm(1,c(2,2))
+     group2[i,] <- rmvnorm(1,c(2,2))
+   }
+ }
R> 
R> hdp_mvnorm <- DirichletProcessHierarchicalMvnormal2(list(group1,group2))
R> hdp_mvnorm <- Fit(hdp_mvnorm, 500)
\end{CodeInput}
After generating the synthetic data we fit a hierarchical multivariate normal Dirichlet process. The resulting clusters in Figure~\ref{fig:hierNormal} are common across both datasets.

\subsection{Posterior Draws and Stick-Breaking Summaries}
The stick-breaking representation in Section \ref{sec:background} provides a convenient way to draw from the posterior random measure associated with a fitted Dirichlet process mixture. In the \pkg{dirichletprocess} package this is exposed through two lower-level helper functions:
\begin{itemize}
\item \code{PosteriorClusters}: draws stick-breaking weights and atoms from the posterior random measure conditional on the current fitted state, and returns them as a list.
\item \code{PosteriorFunction}: draws such a posterior random measure and combines it with the mixture kernel to return a sampled mixture density or mass function.
\end{itemize}
These functions are conditional on the current fitted state of the \code{dirichletprocess} object. They are useful when a workflow requires a single draw from the posterior random measure, as in the point-process example below, but they should not be interpreted as MCMC-averaged summaries. For ordinary fitted objects, posterior means and pointwise credible intervals across retained samples are obtained with \code{PosteriorSummary}.

\subsubsection{Example: Point Process Intensity Estimation}
One practical application of Beta mixture models is the estimation of an inhomogeneous Poisson process intensity functions as in \cite{taddy_mixture_2012}. A Poisson process is a collection of points in space distributed with rate $\lambda$. In the inhomogeneous case, the intensity is dependent on time and as such the number of events in some window $\left[0 , T\right]$ can be written as
\begin{equation*}
N \sim \text{Poisson} (\lambda (t) ).
\end{equation*}
In parametric estimation, a functional form of $\lambda (t)$ would be constructed i.e. $\alpha _0 + \alpha t$ and the parameters $\{\alpha _0, \alpha \}$ would be estimated. However, the accuracy of such a method would be dependent on correctly identifying the parametric form of $\lambda (t)$. With the nonparametric methods that a DPMM provides, such assumptions can be ignored and an intensity function can be built without the need to specify a parametric form. Firstly, we assume that $\lambda (t) = \lambda _0 h(t)$ where $\int _0 ^T h(t) \mathrm{d} t = 1$, i.e. the intensity rate can be decomposed into an amplitude $\lambda _0$ controlling the number of events and a density $h(t)$ controlling the distribution of the events over the window of observation $\left[0 , T\right]$. To infer the value of $\lambda _0$ a conjugate Gamma prior can be used and thus the posterior distribution can be directly sampled.

In this example, we will instead be estimating an intensity rate $\lambda (t)$ with each iteration and using it to update the data. The full model can be written as
\begin{align*}
N & \sim \text{Poisson} ( \lambda (t) ), \\
\lambda (t) & = \lambda _0 h(t), \\
h (t) & = \int k(t \mid \theta) \mathrm{d}F, \\
F & \sim \text{DP} (\alpha , G_0),
\end{align*}
where $k$ and $G_0$ are as per Section \ref{subsec:implemented} for the Beta distribution mixture models. We sample the posterior distribution of $G$ using Key Property 5 (Section \ref{sec:background}) which states that a sample of $G$ can be drawn independently of the data using the stick breaking representation of the data and the model parameters $\theta$.

In this toy model we simulate 500 event times using the intensity function $\lambda (t) = \sin ^2 \frac{t}{50}$. Instead of passing the full data set into the Dirichlet process object, we just use a random sample of 100 of these event times.

\begin{CodeInput}
R> y <- cumsum(runif(1000))
R> intensityShape <- function(x) sin(x/50)^2
R> accept_prob <- intensityShape(y)
R> pts <- sample(y, 500, prob=accept_prob)
\end{CodeInput}
We then fit the Dirichlet process, draw a posterior sample of the intensity function $\hat{\lambda} (t)$ and sample 150 new points from the full data set with probabilities proportional to $\hat{\lambda} (t)$. The Dirichlet process object is then modified with the new data and the process is repeated.

\begin{CodeInput}
R> dp <- DirichletProcessBeta(sample(pts, 100), maxY = max(pts)*1.01,
+ alphaPrior = c(2, 0.01))
R> dp <- Fit(dp, 100, updatePrior = TRUE)
R> 
R> for(i in seq_len(2000)){
+   lambdaHat <- PosteriorFunction(dp)
+   newPts <- sample(pts, 150, prob=lambdaHat(pts))
+   newPts[is.infinite(newPts)] <- 1
+   newPts[is.na(newPts)] <- 0
+   dp <- ChangeObservations(dp, newPts)
+   dp <- Fit(dp, 2, updatePrior = TRUE, progressBar = FALSE)
+ }
\end{CodeInput}
During this loop, \code{PosteriorFunction} is used as a lower-level conditional-on-current-state tool to construct the proposal weights. After the iterative fitting scheme has finished, we summarise the retained states from this scheme using \code{PosteriorSummary} and compare the resulting estimate to the target intensity shape. Since the Beta mixture estimates the normalised density $h(t)$ rather than the amplitude $\lambda_0$, the true curve is scaled to integrate to one over the observation window. The grid is started slightly above zero to avoid numerical boundary issues when evaluating Beta mixture components at the edge of their support.

\begin{CodeInput}
R> xGrid <- seq(0.1, max(pts)*1.01, by=0.1)
R> Tmax <- max(pts)*1.01
R> normConst <- Tmax/2 - 25*sin(Tmax/25)
R> posteriorSummary <- PosteriorSummary(dp, xGrid,
+                                      burnin = 3000,
+                                      thinning = 10,
+                                      level = 0.90)
R> 
R> trueFrame <- data.frame(y=intensityShape(xGrid)/normConst,
+                         x=xGrid)
R> 
R> ggplot() +
+   geom_ribbon(data=posteriorSummary, aes(x=x, ymin=Lower, ymax=Upper),
+               alpha=0.2, fill="red", colour=NA) + #pointwise credible intervals
+   geom_line(data=posteriorSummary, aes(x=x, y=Mean), colour="red") + #mean
+   geom_line(data=trueFrame, aes(x=x, y=y)) #true intensity shape
\end{CodeInput}
\begin{figure}[tb]
\centering
\includegraphics[width=0.45\textwidth]{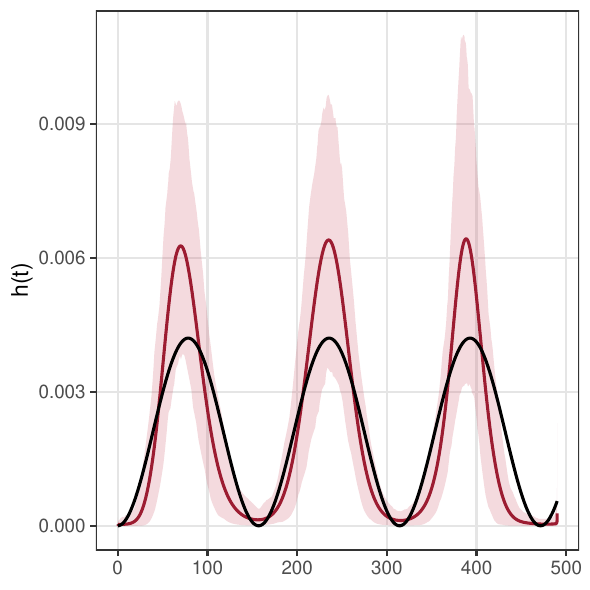}
\caption{Posterior mean and pointwise credible intervals from retained MCMC samples for the normalised intensity shape in the inhomogeneous Poisson process example.}
\label{fig:poissonStick}
\end{figure}
Figure \ref{fig:poissonStick} shows that the main features of the target intensity shape are recovered even though the full dataset is never observed.

\section{Advanced Features}
The material in this section can largely be skipped as long as the user is getting good results from the \pkg{dirichletprocess} package using the default functions and specifications. However some users will require more control over the DP implementation -- perhaps the default base measure hyper parameters are inadequate for a particular data set, or the sampler seeming not to converge due to bad initial default parameter values. Alternatively, the user may wish to use a mixture kernel other than the ones we have included in the package (Normal, Beta, Weibull, etc). In this case, the user will need to know  what is going on under the hood so that they can change the default settings to better suited values, or otherwise modify the internal sampling procedure. We have done our best to ensure that this will usually only require changing a small number of the parameters which control the sampling behaviour, but understanding what needs to be changed (and why) requires some understanding of how the objects are constructed. Note that parts of this section do require an intermediate level of R programming knowledge.

\subsection{Structure of a DP Object: The Gory Details}
This package implements Dirichlet Processes as S3 objects in R. All DPs have the following fields  available which are defined upon construction and do not ever change. When using the DP objects implemented in this package, these will be set by the constructor functions and so can largely be ignored for the default mixture models.

A DP object is defined by its kernel mixing distribution $k(y \mid \theta)$. Each mixing distribution has the following functions and variables associated with its class
\begin{itemize}
\item \code{Likelihood(...)}: a function which specifies the density of the mixture kernel $k(y \mid \theta)$.
\item {\code{PriorDraw(...)}}: a function which returns a random sample of size $n$ from the DP base measure $G_0$. This is used to define $G_0$.
\item \code{g0Priors}: a list of parameters for the base measure $G_0$. Again, this is used to define $G_0$.
\end{itemize}

For a conjugate mixing distribution the posterior distribution of $\theta$ is tractable and can be sampled from directly and the marginal distribution of the data can also be evaluated. Therefore two more functions are needed to complete the specification of a conjugate DP mixture model:
\begin{itemize}
\item {\code{PosteriorDraw(...)}}: a function that returns a sample of size $n$ given data $y$ from the posterior distribution of $\theta$, i.e. a sample from the distribution of $p(\theta \mid y)$.
\item \code{Predictive(...)}: a function that returns the value of the marginal distribution of the data $f(y) = \int k(y , \theta) \mathrm{d} G(\theta)$.
\end{itemize}

With these specified, the \textbf{Fit} function can be used to fit the DP, which carries out the Chinese Restaurant Process sampler using the conjugate Algorithm 4 from \citep{neal_markov_2000}.

For a non-conjugate mixing distribution we can no longer directly sample from the posterior distribution $p(\theta \mid y)$ or calculate the marginal distribution of the data. Instead the Metropolis-Hastings algorithm is used to sample from the distribution $p(\theta \mid y)$. The Metropolis-Hastings algorithm works by generating a candidate parameter $\theta ^{i+1}$ and accepting this candidate value as a sample from the posterior with probability proportional to $\frac{k(y \mid \theta ^{i+1}) p(\theta ^{i+1})}{k(y \mid \theta^i) p(\theta^i)}$. Typically, the candidate parameter is distributed as $\theta _{i+1} \sim N(\theta _i, h^2)$. From this, the non-conjugate mixture model requires 2 additional functions and an extra parameter to be defined.
\begin{itemize}
\item \code{PriorDensity(...)}: a function which evaluates $p(\theta)$ which is the DP base measure $G_0$ for a given $\theta$.
\item \code{mhParameterProposal(...)}: a function that returns a candidate parameter to be evaluated for the Metropolis-Hastings algorithm.
\item \code{mhStepSize}: $h$, the size of the step to make when proposing a new parameter for the Metropolis-Hastings algorithm.
\end{itemize}

With these specified, the \textbf{Fit} function can again be used to fit the DP, which carries out the Chinese Restaurant Process sampler using `Algorithm 8` \citep{neal_markov_2000}.

Once the appropriate mixing distribution is defined we can create a \pkg{dirichletprocess} object which contains the data, the mixing distribution object and the parameter $\alpha$. Then the rest of \pkg{dirichletprocess} class functions are available.

By using the default constructor functions \code{DirichletProcessBeta/Gaussian/Mvnormal/Weibull} the base measure prior parameters are chosen to be non-informative, see Section  \ref{subsec:implemented} for the specific values of the prior parameters.

\subsection{Creating New Dirichlet process Mixture Types} \label{sec:NewMDobj}
The \pkg{dirichletprocess} package currently implements Dirichlet process mixture models using Gaussian, Beta and Weibull kernels. While these kernels should be appropriate for most applications, there will inevitably be times when a user wants to fit a DP model for a kernel which has not been implemented, or otherwise wants to do something complex with a DP which goes beyond the scope of this package. To anticipate this, we have tried to make it easy for users to construct their own mixture models, which can then automatically use the implemented algorithms for fitting a Dirichlet process.

The functions in the package are designed to work on S3 \proglang{R} objects, where each object represents a type of Dirichlet process mixture (e.g Gaussian or Beta). In order to create new types of Dirichlet process mixtures, the user must create a new S3 object type which encapsulates this model and ensure that its specifications correspond to those of the package. If this is done, then all the package functions for re-sampling/prediction/etc should continue to function correctly on the new DP type. This means that the package can hopefully be used for DP applications that we did not consider when writing it, while saving the user from having to write their own functions for re-sampling and fitting.

To illustrate how this works, this section will work through an extended example of how to create a new S3 type which represents a DP mixture model not implemented in the \pkg{dirichletprocess} package. We will explain how the S3 objects are constructed in detail so that the user will be able to create their own.

\subsubsection{Conjugate Mixture}
Suppose we have a particular scenario that requires a Dirichlet process mixture of Poisson distributions. The conjugate prior for the Poisson distribution is the Gamma distribution.

First, we start with the likelihood of the Poisson distribution
\begin{equation*}
k(x \mid \theta) = \frac{\theta ^x \exp(-\theta)}{x!},
\end{equation*}
as there is only one parameter in the Poisson distribution the parameter list $\theta$ is of length 1.
\begin{CodeInput}
Likelihood.poisson <- function(mdobj, x, theta){
  return(as.numeric(dpois(x, theta[[1]])))
}
\end{CodeInput}
Note that the [[1]] part is essential, since parameters are internally represented as lists even when they only have one element.

We then write the random prior draw function which draws a value of $\theta$ from the base measure $G_0$. The conjugate prior to the Poisson distribution is the Gamma distribution
\begin{equation*}
G_0 \sim \text{Gamma} (\alpha _0, \beta _0).
\end{equation*}
\begin{CodeInput}
PriorDraw.poisson <- function(mdobj, n){
  draws <- rgamma(n, mdobj$priorParameters[1], mdobj$priorParameters[2])
  theta <- list(array(draws, dim=c(1,1,n)))
  return(theta)
 }
\end{CodeInput}
The prior parameters $\alpha_0, \beta_0$ are stored in the mixing distribution object \code{mdobj}.

We then write the \code{PosteriorDraw} function to sample from the posterior distribution of $\theta$. Again, as the base measure $G_0$ is conjugate this is a direct sample from the posterior distribution
\begin{equation*}
\theta \mid x \sim \text{Gamma} (\alpha _0 + \sum_{i=1} ^n x_i, \beta_0 + n),
\end{equation*}
using the inbuilt \code{rgamma} function this is trivial.
\begin{CodeInput}
PosteriorDraw.poisson <- function(mdobj, x, n=1){
  priorParameters <- mdobj$priorParameters
  lambda <- rgamma(n, priorParameters[1] + sum(x), priorParameters[2] + nrow(x))
  return(list(array(lambda, dim=c(1,1,n))))
}
\end{CodeInput}

Finally the marginal distribution of the data $f(y)$ can be evaluated as it is a conjugate mixture model and translated into the appropriate \proglang{R} function:
\begin{CodeInput}
Predictive.poisson <- function(mdobj, x){
  priorParameters <- mdobj$priorParameters
  pred <- numeric(length(x))
  for(i in seq_along(x)){
    alphaPost <- priorParameters[1] + x[i]
    betaPost <- priorParameters[2] + 1
    pred[i] <- (priorParameters[2] ^  priorParameters[1]) / gamma(priorParameters[1])
    pred[i] <- pred[i] * gamma(alphaPost) / (betaPost^alphaPost)
    pred[i] <- pred[i] * (1 / prod(factorial(x[i])))
  }
  return(pred)
}
\end{CodeInput}

With these functions written for the Poisson mixture model we now need to use the \code{MixingDistribution} constructor function to create a new object that can be used by the Dirichlet process constructor function, \code{DirichletProcessCreate}.

The constructor function \code{MixingDistribution} creates an object of class \code{distribution}, in this case 'poisson', with prior parameters $\alpha_0, \beta_0 = 1$ and that it is conjugate.
\begin{CodeInput}
R> poisMd <- MixingDistribution(distribution="poisson",
+                              priorParameters = c(1, 1),
+                              conjugate="conjugate")
\end{CodeInput}
This object is now ready to be used in a \pkg{dirichletprocess} object and the appropriate sampling tasks can be carried out. To demonstrate we simulate some test data and fit a Dirichlet process with our new mixing distribution.
\newpage %stop split
\begin{CodeInput}
R> y <- c(rpois(150, 3), rpois(150, 10)) #generate sample data
R> dp <- DirichletProcessCreate(y, poisMd)
R> dp <- Initialise(dp)
R> dp <- Fit(dp, 1000)
R> 
R> posteriorSummary <- PosteriorSummary(dp, 0:20,
+                                      burnin = 500,
+                                      thinning = 5,
+                                      level = 0.90)
R> 
R> trueFrame <- data.frame(x= 0:20,
+                         y= 0.5*dpois(0:20, 3) + 0.5*dpois(0:20, 10))
R> 
R> ggplot() +
+     geom_ribbon(data=posteriorSummary,
+                 aes(x=x, ymin=Lower, ymax=Upper),
+                 colour=NA,
+                 fill="red",
+                 alpha=0.2) + #pointwise credible intervals
+     geom_line(data=posteriorSummary, aes(x=x, y=Mean), colour="red") + #mean
+     geom_line(data=trueFrame, aes(x=x, y=y)) #true
R> 
R> 
\end{CodeInput}
\begin{figure}[tb]
\centering
	\includegraphics[width=0.5\textwidth]{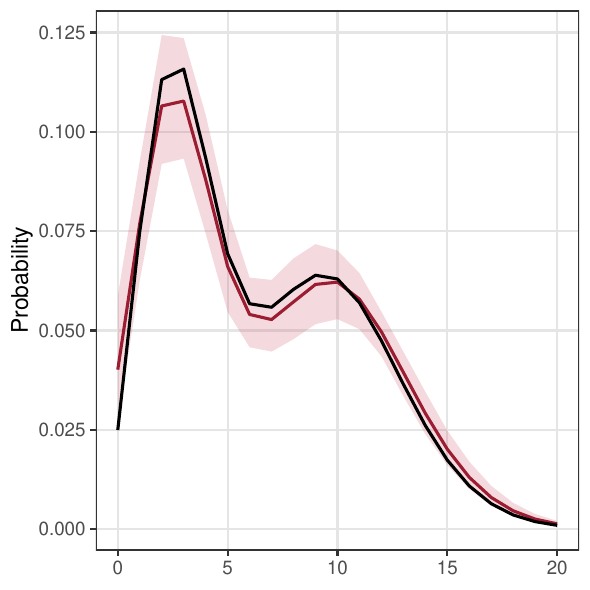}
\caption{The true distribution and the estimated posterior mean and pointwise credible intervals from retained MCMC samples for the Poisson mixture model.}
\label{fig:poissonmixture}
\end{figure}
As Figure \ref{fig:poissonmixture} shows, the fitted mixture closely approximates the true generating distribution. This illustrates how users can create their own mixture models while still relying on the fitting and summary tools provided by the \pkg{dirichletprocess} package.

\subsubsection{Nonconjugate Mixture}
Suppose that a particular application requires a Dirichlet process mixture of Gamma distributions. As the Gamma distribution does not have a conjugate prior distribution additional steps must be taken when creating the necessary functions.

Firstly we must write the likelihood function, as the Gamma distribution has two parameters $\alpha, \beta$ the list $\theta$ will also have two components. The density of the Gamma distribution can be written as
\begin{equation*}
k(y \mid \alpha, \beta) = \frac{\beta ^ \alpha}{\Gamma (\alpha)} y^{\alpha -1} e ^{- \beta y},
\end{equation*}
which can be easily translated using \code{dgamma} in R.
\begin{CodeInput}
Likelihood.gamma <- function(mdobj, x, theta){
  return(as.numeric(dgamma(x, theta[[1]], theta[[2]])))
}
\end{CodeInput}

We now need the function to draw random parameters from the base measure $G_0$. For the parameters of the Gamma distribution we will be using a prior distribution of an Exponential distribution
\begin{align*}
\alpha & \sim \text{Exp} (\alpha _0), \\
\beta & \sim \text{Exp} (\beta _0).
\end{align*}

\begin{CodeInput}
PriorDraw.gamma <- function(mdobj, n=1){
  theta <- list()
  theta[[1]] = array(rexp(n, mdobj$priorParameters[1]), dim=c(1,1, n))
  theta[[2]] = array(rexp(n, mdobj$priorParameters[2]), dim=c(1,1, n))
  return(theta)
}
\end{CodeInput}

Now as we are drawing from the posterior distribution using the Metropolis-Hastings algorithm, we also need a function that calculates the prior density for a given ${\alpha, \beta}$.
\begin{CodeInput}
PriorDensity.gamma <- function(mdobj, theta){
  priorParameters <- mdobj$priorParameters
  thetaDensity <- dexp(theta[[1]], priorParameters[1])
  thetaDensity <- thetaDensity * dexp(theta[[2]], priorParameters[2])
  return(as.numeric(thetaDensity))
}
\end{CodeInput}

Finally, the Metropolis-Hastings algorithm also needs a function that perturbs the parameters to explore the posterior distribution. As for the Gamma distribution the parameters $\alpha, \beta >0$ we must constrain our proposals. This is achieved by taking the absolute value of a standard normal perturbation.

\begin{align*}
\alpha ^{i+1} & = \lvert \alpha ^{i} + h \cdot \eta \lvert ,\\
\eta & \sim N(0, 1), \\
\beta ^{i+1} & = \lvert \beta ^{i} + h \cdot \zeta \lvert, \\
\zeta & \sim N(0, 1). \\
\end{align*}
Again this  is easy to translate into \proglang{R}:
\begin{CodeInput}
MhParameterProposal.gamma <- function(mdobj, oldParams){
  mhStepSize <- mdobj$mhStepSize
  newParams <- oldParams
  newParams[[1]] <- abs(oldParams[[1]] + mhStepSize[1]*rnorm(1))
  newParams[[2]] <- abs(oldParams[[2]] + mhStepSize[2]*rnorm(1))
  return(newParams)
}
\end{CodeInput}

We can now construct our mixing distribution object using the constructor function \code{MixingDistribution}. The arguments of this function specify the prior parameters $\alpha _0, \beta _0$ and set the scale $h$ at which the new parameter proposals are made using the parameter \code{mhStepSize}.
\begin{CodeInput}
gammaMd <- MixingDistribution(distribution = "gamma",
                              priorParameters = c(0.1, 0.1),
                              conjugate = "nonconjugate",
                              mhStepSize = c(0.1, 0.1))
\end{CodeInput}

The dirichletprocess object can now be created and fit to some test data. As it is a new type of mixture, it must be initialised.
\begin{CodeInput}
R> y <- c(rgamma(100, 2, 4), rgamma(100, 6, 3)) #generate sample data
R> dp <- DirichletProcessCreate(y, gammaMd)
R> dp <- Initialise(dp)
R> dp <- Fit(dp, 1000)
R> 
R> posteriorSummary <- PosteriorSummary(dp, ppoints(100)*6,
+                                      burnin = 500,
+                                      thinning = 5,
+                                      level = 0.90)
R> 
R> trueFrame <- data.frame(x=ppoints(100)*6,
+                           y= 0.5*dgamma(ppoints(100)*6, 2, 4) +
+                           0.5*dgamma(ppoints(100)*6, 6, 3))
R> 
R> ggplot() +
+   geom_ribbon(data=posteriorSummary,
+               aes(x=x, ymin=Lower, ymax=Upper),
+               colour=NA, fill="red", alpha=0.2) +
+   geom_line(data=posteriorSummary, aes(x=x, y=Mean), colour="red") +
+   geom_line(data=trueFrame, aes(x=x, y=y))
\end{CodeInput}
\begin{figure}[tb]
\centering
	\includegraphics[width=0.45\textwidth]{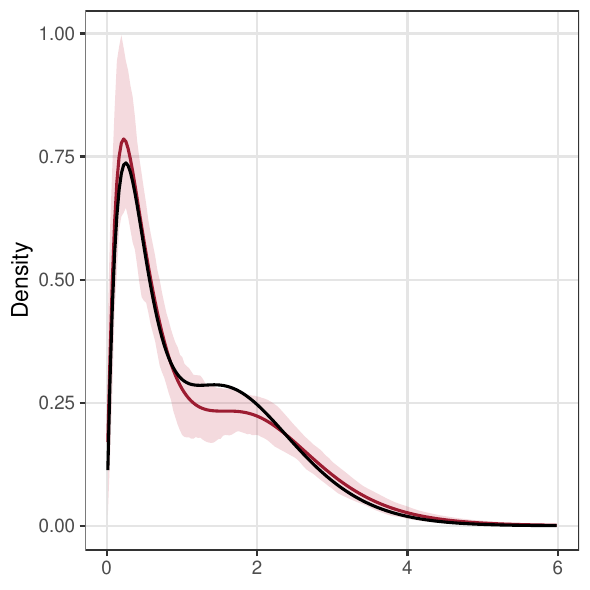}
\caption{The results of implementing the new Gamma mixture model, with posterior mean and pointwise credible intervals from retained MCMC samples.}
\label{fig:gammamixture}
\end{figure}
Figure \ref{fig:gammamixture} shows that the fitted mixture closely approximates the true generating distribution.

\subsection{Extended Example - Working with Censored Observations}
%use the excellent example from my code
The following example is intended to illustrate how simple the \pkg{dirichletprocess} package makes it for the user to extend Dirichlet process mixture modeling to situations which are not directly supported by the package, in a less trivial context than the above density estimation examples.

Survival analysis is an area of statistics concerned with analysing the duration of time before an event happens such as a failure in a mechanical system or a death in a biological context. We are concerned with constructing a survival function which as the name indicates shows how the probability of not experiencing the event changes with time. Survival data is often generated by observational studies which result in censoring. In the context of medical statistics, censoring occurs due to finite time periods of the studies. When analysing the effects of medical treatments patients events can be censored for a variety of reasons. This is a missing data problem as we no longer know the exact time at which an event occurred, just that it occurred before or after a specific time. Right censoring is when a patient is yet to be affected by the event after a study ends. Left censoring is when it is not known exactly when the event occurred, just that it occurred before the study started. To deal with this censored information we must adapt our likelihoods.

One  nonparametric approach for modeling such data  is to use a Dirichlet process mixture of Weibull distributions. The \pkg{dirichletprocess} package does not directly support the analysis of censored data -- as stated throughout, the purpose of the package is not to provide the user with a small number of functions for solving predefined problems, but instead to make it easy to use Dirichlet process mixtures in a wide variety of contexts. As such, it is very simple for the user to extend the functionality of the package to allow for censoring.

For the data we replicate the work of \cite{kottas_nonparametric_2006} and use leukaemia remission times taken from \cite{lawless_statistical_2011}. This dataset contains two groups of censored observations and we wish to investigate the impact two different treatments have on leukaemia remission times:

\begin{CodeInput}
R>  data_a <- c(1, 3 ,3, 6, 7, 7, 10, 12, 14, 15, 
+      18 ,19, 22 ,26 , 28 , 29 ,34, 40, 48 ,49)
R>  data_b <- c(1, 1, 2, 2,3,4,5,8,8,9,11,12,14,16,18,21,27,31,38, 44)
R>  scale_factor <- max(data_a,data_b)
R>  data_a <- 1 + (data_a / scale_factor)
R>  data_a <- cbind(data_a, c(0,0,0,0,0,0,0,0,0,0,0,0,0,0,1,0,0,0,1,1))
R>  data_b <- 1 + (data_b / scale_factor)
R>  data_b <- cbind(data_b, c(0,0,0,0,0,0,0,0,0,0,0,0,0,0,0,0,1,0,1,0))
\end{CodeInput}

Since the Weibull hyperparameters are scale-dependent, we rescale the remission times to a common interval before fitting so that a fixed set of hyperparameters can be used across examples. The transformed times are shifted away from zero to avoid numerical instability in the Weibull likelihood. We next fit a censored Weibull DP model to each of the data sets and compare the survival functions. The full model can be written as
\begin{align*}
y _i & \sim \text{Weibull} ( y_i \mid a_i , b_i ), \\
a_i , b_i & \sim G, \\
G & \sim  \text{DP} (\alpha , G_0),
\end{align*}
where the Weibull density and $G_0$ follow the form shown in Section \ref{subsec:implemented}.

Censored data can come in the form of two columns - the time it takes for the event to occur and an indicator variable; 1 for a right censored observation, 0 otherwise. Therefore the density of the Weibull distribution can be written as
\begin{align*}
k(y \mid a , b) & = \frac{a}{b} y ^{a-1}  \exp \left( -  \frac{y^a}{b}  \right) \quad \text{for uncensored}, \\
k(y \mid a , b) & =  \exp \left( -\frac{y^a}{b} \right) \quad \text{for censored}.
\end{align*}
Therefore we must translate this likelihood into the appropriate function for the \pkg{dirichletprocess} package.

\begin{CodeInput}
Likelihood.weibullcens <- function(mdobj, x, theta) {
  a <- rep_len(as.numeric(theta[[1]]), max(length(theta[[1]]), length(theta[[2]]), NROW(x)))
  b <- rep_len(as.numeric(theta[[2]]), length(a))
  t <- rep_len(x[, 1], length(a))
  cens <- rep_len(x[, 2], length(a))

  y <- rep(0, length(a))
  ok <- is.finite(a) & is.finite(b) & a > 0 & b > 0

  if (any(ok)) {
    log_y <- log(a[ok]) - log(b[ok]) + (a[ok] - 1) * log(t[ok]) - t[ok]^a[ok] / b[ok]
    log_y[cens[ok] == 1] <- -t[ok][cens[ok] == 1]^a[ok][cens[ok] == 1] / b[ok][cens[ok] == 1]
    y[ok] <- exp(log_y)
  }

  y[!is.finite(y)] <- 0
  y
}
\end{CodeInput}

We now wish to create two different \code{mixingDistribution} objects for each of the data sets. Again we use the \texttt{MixingDistribution} constructor function and assign the resulting
object a custom S3 class. The custom class \texttt{weibullcens} is placed first, so that
method dispatch uses the censored-Weibull likelihood where available, while still inheriting
the more general behaviour associated with the \texttt{weibull} and \texttt{nonconjugate}
classes.

\begin{CodeInput}
R> mdobjA <- MixingDistribution("weibullcens",
+                              c(1,2,0.5), "nonconjugate",
+                              mhStepSize=c(0.11,0.11),
+                              hyperPriorParameters=c(2.222, 2, 1, 0.05))
R> mdobjB <- MixingDistribution("weibullcens",
+                              c(1,2,0.5), "nonconjugate",
+                              mhStepSize=c(0.11,0.11),
+                              hyperPriorParameters=c(2.069, 2, 1, 0.08))
R> 
R> class(mdobjA) <- c("weibullcens", "weibull", "nonconjugate")
R> class(mdobjB) <- c("weibullcens", "weibull", "nonconjugate")
\end{CodeInput}
We can easily use this modified mixture model for our censored data. The sampling is then carried out as normal with no other changes needed. The default functions available for the Weibull mixture model are applied to our custom dirichletprocess object.
\begin{CodeInput}
R> dpA <- DirichletProcessCreate(data_a, mdobjA, c(2, 0.9))
R> dpA <- Initialise(dpA)
R> 
R> dpB <- DirichletProcessCreate(data_b, mdobjB, c(2, 0.9))
R> dpB <- Initialise(dpB)
R> 
R> dpA <- Fit(dpA, 500, updatePrior = TRUE)
R> dpB <- Fit(dpB, 500, updatePrior = TRUE)
\end{CodeInput}
Using the fitted values we can extract chain-averaged point estimates of the density and survival functions using the weights and cluster parameters. The survival function is calculated as $S(y) = \exp(- \frac{y ^a}{b})$.
\begin{figure}
\centering
\includegraphics[width=0.8\textwidth]{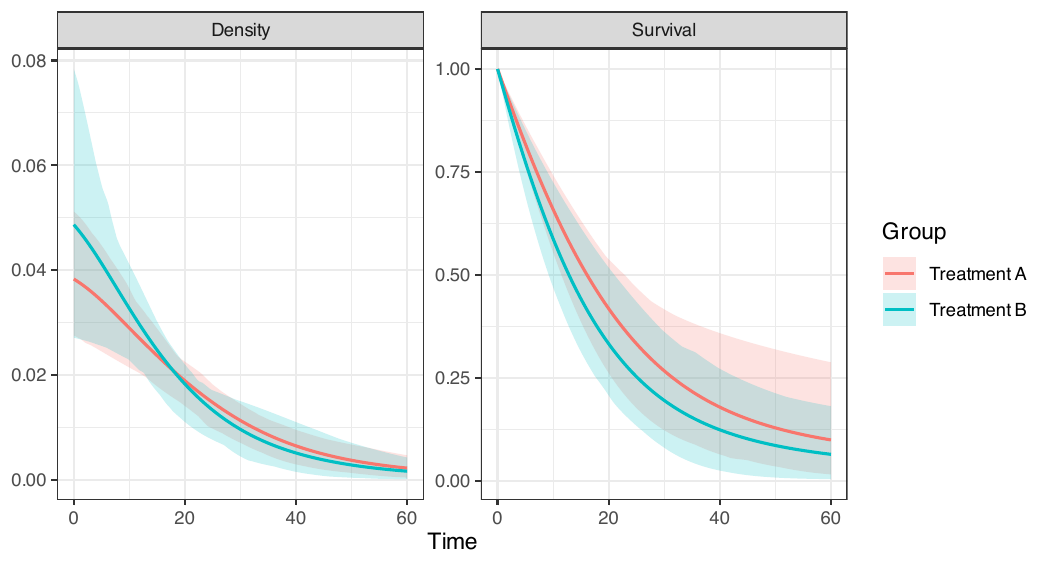}
\caption{Chain-averaged point estimates for the survival and density functions of the two treatments.}
\label{fig:weibull_cens}
\end{figure}
The resulting density and survival estimates are shown in Figure \ref{fig:weibull_cens}, with the data rescaled to their original values. The results are consistent with the qualitative treatment comparison in \cite{kottas_nonparametric_2006}.

To fully understand what has happened here, it is vital to understand that the DP is defined by its base likelihood and $G_0$ distribution. In creating a new mixing distribution with custom class \texttt{weibullcens}, we introduce a new likelihood while reusing the more general functionality already available for Weibull and
non-conjugate Dirichlet process mixtures. This allows users to extend existing mixture models
with relatively small modifications, without having to re-implement the full fitting machinery.

\subsection{Resampling Component Indexes and Parameters}
When calling the \code{Fit} function on a DP object the component indexes and parameters are resampled following Algorithm 4 for the conjugate case and Algorithm 8 for the non-conjugate case from \cite{neal_markov_2000}. For both types of DP mixture the two functions that do the majority of the work are \code{ClusterComponentUpdate} and \code{ClusterParameterUpdate}.

In a conjugate DPMM new component indexes and new cluster parameters are drawn directly from the predictive and posterior distributions making the algorithm very efficient. In such case, the only option available to users is to change the prior parameters of the base distribution $G_0$. Ensuring that the base distribution is correctly parameterised with sensible values for the underlying data will provide optimal performance for the fitting algorithm.

However, in a non-conjugate case new cluster components are proposed from the chosen prior distribution and new cluster parameters are sampled using the Metropolis-Hastings algorithm to obtain a posterior sample. By using the Metropolis-Hastings algorithm, the parameters in question are proposed using a random walk but constrained to the particular support of the parameter. For example, the parameters in a Weibull distribution are strictly positive, therefore the random walk is restricted to fall on the positive real line. An ill-proposed prior distribution can severely affect the convergence of the fitting process. The parameter \code{mhStepSize} in the constructor function for a non-conjugate mixture controls the scale of new parameter proposals for the random walk. When creating a new DP object, the constructor function has a flag \code{verbose} that outputs an estimated acceptance ratio, for optimal performance of the Metropolis-Hastings algorithm this value should be around $0.234$ \citep{gelman_efficient_1996}. Therefore the user should adjust \code{mhStepSize} to reach this value. As with the conjugate case, care must be taken to ensure that the base measure is well suited for the data.

\subsubsection{Overriding Default Behaviour}
For both conjugate and non-conjugate mixture models, the user can write their own \\ \code{ClusterComponentUpdate} and \code{ClusterParameterUpdate} functions to override the default behaviour. The user can still benefit from the other S3 methods and structures implemented in \pkg{dirichletprocess} but with their custom sampling schemes.

For the non-conjugate mixture models there is a further option available to change the component index and parameter re-sampling. In Algorithm 8 of \cite{neal_markov_2000} each datapoint can form a new cluster with parameter drawn from the base measure, these proposals are called `auxiliary' variables and $m$ are drawn for each data point. By default $m=3$. However this can be changed in the \code{Initialise(dp, ...,m=m)} function. Using more auxiliary variables can lead to more changes in the component indexes and greater exploration of the base measure but at the cost of computational time.

\subsection{Resampling the Base Measure, $G_0$}
It is helpful that the user knows how to best set the parameters of the base measure to correctly represent the underlying data. However, whilst desirable this is not always practical. In which case the package offers functionality to use hyper-prior parameters on $G_0$ and update them with each iteration.

For the mixing distributions that allow for re-sampling of the base measure, it is simple to include the flag \code{Fit(dp,...,updatePrior=TRUE)}. At each fitting iteration the base measure with variable parameters will be updated based on the current cluster parameters. For details on the exact specification of the hyper-prior distributions for each implemented mixture kernel see Section \ref{subsec:implemented}.  If a user wishes to change the default prior on the hyper parameters then it is as simple as changing the \code{PriorParametersUpdate} function for the mixing distribution object.

%\subsection{Resampling $\alpha$ }
%\subsubsection{Overriding Default Behavior}

%\subsection{Working with DP objects - Prediction, Stick Breaking Representation, Modifying the Observations Being Analysed, etc}

\subsection{Component Prediction}
Given a fitted DP object and some new data $\hat{y}$ to predict the cluster labels of the new data we use the command \code{ClusterLabelPredict}. Using the appropriate algorithm for a conjugate or non-conjugate mixture model the cluster label probabilities are calculated from the new data $\hat{y}$, these probabilities are then sampled once to obtain a cluster label. It is these cluster labels that are returned with the appropriate cluster parameters.

We refer back to our example in Section \ref{subsec:clusteranalysis} where we used a Dirichlet process to cluster the \code{faithful} dataset. In this example, we will withhold the last five entries of the data as the prediction set and use \code{ClusterLabelPredict} to estimate their cluster allocation.

\begin{CodeInput}
R> faithfulTrans <- scale(faithful)
R> trainIndex <- 1:(nrow(faithfulTrans)-5)
R> 
R> dp <-  DirichletProcessMvnormal(faithfulTrans[trainIndex, ])
R> dp <- Fit(dp, 1000)
R> 
R> labelPred <- ClusterLabelPredict(dp, faithfulTrans[-trainIndex, ])
\end{CodeInput}
The function \code{ClusterLabelPredict} works by calculating and sampling the clustering label from the probability that each test point $\hat{y} _j$ belongs to each cluster $\theta _i$ or should form its own cluster proportional to $\alpha$
\begin{align*}
p(i) & \propto n_i k(\hat{y} _j \mid \theta _i), \\
p(i=\text{new}) & \propto \alpha \int k(\hat{y} _j , \theta _i) \mathrm{d} G_0.
\end{align*}
The function returns a list with multiple entries:
\begin{itemize}
\item The predicted cluster labels for the data under \code{labelPred$componentIndexes}.
\item The cluster parameters assigned to the predicted data points in case a new cluster is predicted \code{labelPred$clusterParameters}.
\item The new number of data points per cluster \code{labelPred$pointsPerCluster}.
\item The total number of clusters are also returned \code{labelPred$numLabels} as this can change with each prediction.
\end{itemize}

We can then use the results of this function to form a dataframe and plot the results.

\begin{CodeInput}
R> faithfulTrainPlot <- data.frame(faithful[trainIndex, ],
+                                 Label=dp$clusterLabels)
R> faithfulTestPlot <- data.frame(faithful[-trainIndex, ],
+                                Label=labelPred$componentIndexes)
R> 
R> ggplot() +
+     geom_point(data=faithfulTrainPlot,
+                aes(x=eruptions,
+                    y=waiting,
+                    colour=as.factor(Label)),
+                size=1) +
+     geom_point(data=faithfulTestPlot,
+                aes(x=eruptions,
+                    y=waiting,
+                    colour=as.factor(Label)),
+                shape=17, size=5) +
+   guides(colour='none')
\end{CodeInput}
\begin{figure}[tb]
\centering
	\includegraphics[width=0.5\textwidth]{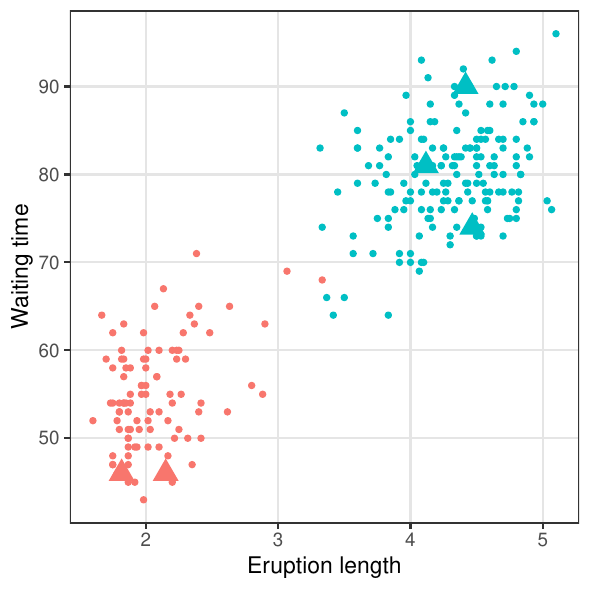}
\caption{The predicted labels of the last 5 entries of the \code{faithful} dataset against the training data. The predicted values are indicated by a solid colour and triangle shapes.}
\label{fig:faithfulpred}
\end{figure}
Figure \ref{fig:faithfulpred} shows the test data being correctly identified with the appropriate cluster.

\section*{Acknowledgements}
We thank Kees Mulder for his contributions. We thank Gianluca Baio and Federico Ricciardi for their helpful comments in the development of this package and vignette.

\bibliography{dirichletprocess}

\end{document}